\renewcommand{\vec}[1]{\mbox{\boldmath $\displaystyle #1$}} %boldface
\newcommand{\grad}{\vec{\nabla}} %gradient
\newcommand{\ddt}[1]{\frac{\partial #1}{\partial t}} %partial time
\newcommand{\DDt}[1]{\frac{d #1}{dt}} %total time derivative
\newcommand{\ddy}[1]{\frac{\partial #1}{\partial y}} %partial derivative
\newcommand{\mdot}{\dot{m}}     %local accretion rate
\newcommand{\medd}{\ \mdot_{\rm Edd}} %local Eddington accretion rate
\newcommand{\delab}{\nabla_{\rm\!ad}} %adiabatic gradient
\documentstyle[aaspp4,epsfig,flushrt]{article}

\begin{document}

\title{The rp Process Ashes from Stable Nuclear Burning on an Accreting Neutron Star}

\author{Hendrik Schatz\footnote{
Current address: Gesellschaft f{\"u}r Schwerionenforschung, Darmstadt,
Germany}, Lars Bildsten and Andrew Cumming}
\affil{Department of Physics and Department of Astronomy\\ 366 LeConte
Hall, University of California, Berkeley, CA 94720 \\ email:
h.schatz@gsi.de, bildsten@fire.berkeley.edu,
cumming@fire.berkeley.edu}

\author{Michael Wiescher} 
\affil{Department of Physics, University of Notre Dame, Notre Dame, IN
46556\\ email:  wiescher.1@nd.edu}

\begin{abstract}
The temperature and nuclear composition of the crust and ocean of an
accreting neutron star depend on the mix of material (the ashes) that
is produced at lower densities by fusion of the accreting hydrogen and
helium. The hydrogen/helium burning is thermally stable at high
accretion rates, a situation encountered in weakly magnetic ($B\ll
10^{11} \ {\rm G}$) neutron stars accreting at rates $\dot M > 10^{-8}
{M_\odot
\ {\rm yr^{-1}}}$ and in most accreting X-ray pulsars, where the
focusing of matter onto the magnetic poles results in local accretion
rates high enough for stable burning. For a neutron star accreting at
these high rates, we calculate the steady state burning of hydrogen
and helium in the upper atmosphere ($\rho < 2\times 10^6 \ {\rm g\
cm^{-3}}$), where $T\approx (5$--$15)\times 10^8 \ {\rm K}$. Since the
breakout from the ``hot'' CNO cycle occurs at a temperature comparable
to that of stable helium burning ($T\gtrsim 5\times 10^8\ {\rm K}$),
the hydrogen is always burned via the rapid proton capture (rp)
process of Wallace and Woosley.

The rp process makes nuclei far beyond the iron group, always leading
to a mixture of elements with masses $A\sim 60$--$100$. The average
nuclear mass of the ashes is set by the extent of helium burning via
($\alpha$,p) reactions, and, because these reactions are temperature
sensitive, depends on the local accretion rate. Nuclear statistical
equilibrium, leading to a composition of mostly iron, occurs only for
very high local accretion rates in excess of 50 times the Eddington
rate.

We briefly discuss the consequences of our results for the properties
of the neutron star. The wide range of nuclei made at a fixed
accretion rate and the sensitivity of the ash composition to the local
accretion rate makes it inevitable that accreting neutron stars have
an ocean and crust made up of a large variety of nuclei. This has
repercussions for the thermal, electrical and structural properties of
the neutron star crust. A crustal lattice as impure as implied by our
results will have the conductivity throughout most of its mass set by
impurity scattering, allowing for more rapid Ohmic diffusion of
magnetic fields than previously estimated for mono-nuclear mixes.

\end{abstract}


\begin{center}
{\bf To Appear in The Astrophysical Journal}
\end{center}

\newpage

%------------------------ INTRODUCTION ------------------------------------

\section{Introduction} 

Neutron stars in mass-transferring binaries accrete hydrogen and
helium rich material from their companions at rates ranging from
$10^{-11}$--$10^{-8} M_\odot \ {\rm yr^{-1}}$. This matter undergoes
thermonuclear fusion within hours to days of reaching the neutron star
surface, releasing $\approx 5 \ {\rm MeV}$ per nucleon for solar
abundances. The nuclear burning is thermally unstable on weakly
magnetic neutron stars ($B\ll 10^{11} \ {\rm G}$) accreting at $\dot M
< 10^{-8} M_\odot \ {\rm yr^{-1}}$ and produces energetic ($\sim
10^{39}\ {\rm erg}$) Type I X-ray bursts when $\dot M < 10^{-9}
M_\odot \ {\rm yr^{-1}}$ (see Lewin, van Paradijs \& Taam 1995 and
Bildsten 1998b for recent reviews; observationally, the nature of the
time dependent burning in the regime $10^{-9} M_\odot \ {\rm yr^{-1}}
< \dot M < 10^{-8} M_\odot \ {\rm yr^{-1}}$ is still not
understood). The composition of the ashes from the unstable burning is
still uncertain, but most certainly consists of heavy nuclei,
potentially beyond the iron group (Hanawa, Sugimoto \& Hashimoto 1983;
Wallace \& Woosley 1984; Schatz et al. 1997; Schatz et al. 1998, 
Koike et al. 1999).

In this paper, we calculate for the first time the mix of nuclei made
during {\it thermally stable} hydrogen/helium burning in the upper
atmosphere of an accreting neutron star. This is appropriate for most
accreting X-ray pulsars, where the local accretion rate is high enough
for stable burning on the magnetic polar cap (Joss \& Li 1980;
Bildsten \& Brown 1997), and for the bright ``Z'' sources of Hasinger
and van der Klis (1989) that are not regular Type I bursters and
accrete globally at $\dot M\sim 10^{-8} M_\odot \ {\rm yr^{-1}}$.
Whereas accurately calculating the ashes from unstable burning
requires time-dependent, three-dimensional modelling of the ignition
and propagation of the flame, stable burning is time-independent and
therefore calculation of the ashes is much more straightforward.

The ashes are forced by accretion into the neutron star ocean and
crust, replacing what was there at birth. The ensuing electron
captures, neutron emissions and pycnonuclear reactions release energy
locally and drive the matter neutron-rich (Bisnovatyi-Kogan \&
Chechetkin 1979; Sato 1979; Haensel \& Zdunik 1990; Blaes
et. al. 1990; Bildsten 1998a; Brown, Bildsten \& Rutledge 1998). Thus
the composition of an accreted or accreting neutron star ocean and
crust is very different from the primordial one (see Pethick and
Ravenhall 1995 for a recent discussion of the primordial crust) and
depends critically on the range and type of nuclei made during the
H/He burning in the upper atmosphere.  Previous studies of accreting
neutron star crusts presumed that iron is the sole product of nuclear
burning in the upper atmosphere and is the only nucleus entering the
neutron star crust. We find that the hydrogen burning on these rapidly
accreting neutron stars is mostly via the rp process of Wallace and
Woosley (1981), resulting in a complicated mix of elements, {\it
nearly all much heavier than iron. }

The thin outer crust of a neutron star (before neutron drip at
$\rho\approx (4$--$6)\times 10^{11} \ {\rm g \ cm^{-3}}$) is replaced
by accretion of only $\approx 10^{-4}M_\odot$ of material. Thus all of
the neutron stars in low-mass X-ray binaries should have an accreted
outer crust. Indeed these objects accrete enough material to
unambiguously replace their {\it whole} crust (down to the crust/core
interface), which is typically a few percent of the total stellar mass
(Ravenhall \& Pethick 1994).  The X-ray pulsars typically accrete
from the wind of a massive companion at a rate $\dot M\sim
10^{-11}$--$10^{-10} M_\odot \ {\rm yr^{-1}}$ (see Bildsten et
al. 1997 for a recent overview) and so are capable of replacing their
outer crusts.  A few of these objects (SMC X-1, LMC X-4, Her X-1) are
accreting at high enough rates or for long enough times (as inferred
by the $2.2\ M_\odot$ mass companion to Her X-1) to replace their
whole crust.

Knowledge of the composition and thermal properties of an accreting
neutron star's ocean and crust is important for many studies. For
example, the temperature and composition of the crust affects the
thermal conductivity and the Ohmic diffusion time there (see Brown \&
Bildsten 1998 for a recent application/overview of this problem). The
crustal composition determines the amount of heat deposited directly
in the crust (Haensel \& Zdunik 1990; Miralda-Escude et al. 1990) and
the rate of neutrino emission from the crust (Haensel, Kaminker \&
Yakovlev 1996), both of which are important for finding the
equilibrium core temperature of an accreting neutron star (Fujimoto et
al. 1984; Brown \& Bildsten 1998). The predicted frequencies of the
so-far unobserved ocean g-modes depend directly on the average nuclear
mass (Bildsten \& Cutler 1995; Bildsten \& Cumming 1998).

We begin in \S 2 with an introduction to the basic equations we solve
and a summary of the input microphysics. In \S 3, we discuss the high
accretion rate burning regime and explain why the burning is thermally
stable at high accretion rates and why the helium ignites in a
hydrogen-rich environment, thus providing an excellent site for the rp
process. In \S 4, we explain the overall thermal and compositional
structure of the burning layer and provide convenient analytic
expressions for the temperature and depth of the burning. Section 5
contains an in-depth discussion of the nature of the rp process. We
explain how the burning depends on the local accretion rate, with
particular emphasis on the important role of the $\alpha$p process in
determining the final average mass of the nuclei. We conclude in \S 6
with a summary of our work and some speculations about how these
results will impact studies of the neutron star crust and ocean. In
the Appendix, we describe how we calculate the radiative and
conductive opacities for the complex mixtures produced by the
hydrogen/helium burning.

%------------------------------ SECTION TWO -------------------------------

\section{Basic Equations and Microphysics}\label{SS}

The plane-parallel nature of the neutron star atmosphere means
that the physics of nuclear burning depends on the accretion rate per
unit area, $\dot m$ (Fujimoto, Hanawa \& Miyaji 1981, hereafter FHM).
This parameter determines the local behavior on the neutron star and
need not be the same everywhere, especially on accreting X-ray
pulsars. The local zero metallicity Eddington rate (the accretion rate
at which the outgoing radiation exerts a force comparable to gravity)
is
\begin{equation} 
\dot m_{\rm Edd}={{2 m_p c}\over {(1+X) R \sigma_{\rm
Th}}}={8.8\times 10^4 \ {\rm g \ cm^{-2} \ s^{-1}}}
\left({1.71 \over {1+X}}\right)
\left({10 \ {\rm km} \over R}\right),
\end{equation} 
where $\sigma_{\rm Th}$ is the Thomson scattering cross-section, $m_p$
is the proton mass, $c$ is the speed of light, $X$ is the hydrogen
mass fraction, and $R$ is the stellar radius. In this paper, we use
the Eddington accretion rate for solar composition ($X=0.71$) and
$R=10$~km, $\dot m_{\rm Edd}=8.8\times 10^4\ {\rm g \ cm^{-2} \
s^{-1}}$, as our basic unit for the local accretion rate.

Steady accretion modifies the equations of particle continuity and
entropy for a hydrostatic settling atmosphere (FHM; Brown \& Bildsten
1998). We write these equations in a coordinate system of fixed
pressure $P$, so that the accreted matter flows through the
coordinates as it is compressed by accretion of fresh material from
above. Hydrostatic balance (the ram pressure of the accretion flow is
negligible) yields $P=gy$, where the column depth $y$ (in g cm$^{-2}$)
is defined by $dy=-\rho\,dz$, and $g\approx GM/R^2$ (we neglect
general relativistic corrections). The continuity equation for an
element $i$ (with number density $n_i$) is
\begin{equation}
\ddt{n_i} + \grad \cdot  (n_i\vec{v}) = \sum r,
\end{equation}
where $\sum r$ is the summed rates of particle creation and
destruction processes.  For these high accretion rates, there is not
time for differential settling and all elements co-move downward at
the speed needed to satisfy mass continuity, $v=\dot m/\rho$ (Wallace,
Woosley \& Weaver 1982; Bildsten, Salpeter \& Wasserman 1993). We
define a mass fraction $X_i\equiv \rho_i/\rho= A_i m_p n_i/ \rho $
where $A_i$ is the baryon number of species $i$ and expand the
continuity equation to obtain
\begin{equation}\label{eq:contin}
\ddt{X_i} + \dot{m} \ddy{X_i} = \frac{A_i m_p\sum r}{\rho}. 
\end{equation}
The equation for the entropy is 
\begin{equation} \label{eq:entropy}
T\DDt{s} = -\frac{1}{\rho}{\grad \cdot \vec{F}} + \epsilon,
\end{equation}
where $\epsilon$ is the energy release rate from nuclear burning and
$\vec{F}$ is the heat flux. We write the entropy as $Tds=C_p
T(dT/T-\delab dP/P)$, where $\delab=d\ln T/d\ln P$ for an adiabatic
change and $C_p$ is the specific heat at constant pressure. Since the
temperature can depend on both time and pressure, we find
\begin{equation}\label{eq:entrop}
\ddy{F} + \epsilon=C_p\left(\ddt{T}+\dot{m}\ddy{T}\right)-\frac{C_pT\mdot}{y}\delab.
\end{equation}
Equations (\ref{eq:contin}) and (\ref{eq:entrop}) describe the
hydrostatic evolution of the neutron star atmosphere while constantly
accreting. In our steady-state calculations, we neglect the
time-dependent terms in these equations.

We solve the continuity and entropy equations in connection with a
nuclear reaction network to determine the nuclear energy generation
rate $\epsilon$ in equation (\ref{eq:entrop}) and the nuclear
abundances $Y_i\equiv X_i/A_i$.  The nuclear reaction network used
here is described in detail in Schatz et al. (1998) and the references
therein
(see also Herndl et al. 1995, Van Wormer et al.
1994, and Wiescher et al. 1986 for discussions of the reaction rates).
The network includes 631 nuclei between hydrogen and
$^{100}$Sn covering the range from stability to the proton drip
line. The types of reactions considered are all proton, neutron and
$\alpha$ induced reactions as well as photodisintegration, $\beta^+$
decay and electron capture. For most nuclei in the range $Z
\le 32$ and $A \le 60$ the temperature and density dependent weak interaction
rates are taken from the compilation of Fuller, Fowler \& Newman
(1980, 1982a, 1982b). These weak interaction rate data include energy
losses due to emission of neutrinos in electron captures and $\beta^+$
decays. Since similar data for heavier nuclei are not available, the
$\beta$-decay rates of nuclei with $Z>32$ were approximated using
temperature and density independent earth rates (for a justification
see Schatz et al. 1998). Energy loss via neutrinos is neglected for
these nuclei. Electron captures do not play an important role for the
density regime in which we work.  Electron screening is treated
according to Graboske et al. (1973) and Itoh et al. (1979).

For the equation of state, we assume the ions behave as an ideal gas
(Coulomb corrections to the ion equation of state are unimportant at
the column depths of interest to us) and we use the interpolation
formulae of Paczynski (1983) to account for the partially degenerate
electrons. The heat flux through the atmosphere is given by
\begin{equation}\label{eq:heat}
F=-{c\over 3\kappa \rho}{d\over dz} aT^4
={c\over 3\kappa } {d\over dy} aT^4,
\end{equation}
where $a$ is the radiation constant. The opacity $\kappa$ is set by
electron scattering, free-free absorption and conduction. In the
Appendix, we discuss in detail how we calculate each of these
contributions for the complex mixtures produced by the
hydrogen/helium burning.

We assume an initial solar composition (Anders and Grevesse 1989) of
the accreted material and a neutron star with $M=1.4~M_\odot$ and
$R=10~{\rm km}$. The influence of these parameters on our results is
discussed later. The only boundary condition that has to be chosen is
the radiation flux exiting the top of the atmosphere. For these
steady-state models at high accretion rates, the flux is dominated by
the nuclear energy release from the conversion of hydrogen and helium
to heavy elements ($\approx$ 5 MeV per nucleon). There is also energy
release from gravitational settling and energy generated in the crust
due to electron captures, neutron emission and pycnonuclear reactions
(see Brown \& Bildsten 1998 for a discussion of this issue). However,
these contributions to the flux are small ($\lesssim 10\%$) compared
to the nuclear energy release, and we do not include them here. In
particular, we drop the terms proportional to $C_p$ on the right-hand
side of equation (\ref{eq:entrop}). Our approach is to guess the flux
at the top of the atmosphere, integrate to the base of the burning
layer and then compare the flux generated by the nuclear burning to
our initial estimate. We find that this procedure converges to give
the correct upper boundary condition in only a few iterations.

%------------------------------ SECTION THREE -----------------------------

\section{The Nature of Hydrogen and Helium Burning at High Accretion Rates}
\label{Burn}

For the high accretion rates of interest to us, the temperature at the
depth where burning occurs is always in excess of $8 \times 10^7$ K,
so the hydrogen burns either via the hot CNO cycle or, at the higher
temperatures at later stages, via the rp process.  During the hot CNO
cycle the timescale for proton captures is shorter than that for
subsequent $\beta$ decays. The time to go around the catalytic loop is
set by the $\beta$ decay lifetimes of $^{14}$O ($t_{1/2}=70.6$~s) and
$^{15}$O ($t_{1/2}=122.2$~s) and is {\it temperature independent}.
The $\beta$-decays fix the hydrogen burning rate at (Hoyle \& Fowler
1965)
\begin{equation} \label{eq:cno_burn}
\epsilon_{\rm CNO}=5.8\times 10^{15} Z_{\rm CNO} {\rm \ ergs  \ g^{-1} \
s^{-1}},
\end{equation} 
where $Z_{\rm CNO}$ is the mass fraction of CNO in the accumulating
matter. The timescale to consume all of the hydrogen is then $\approx
790\,{\rm s}/Z_{\rm CNO}$ or about one day for solar metallicity. For
high accretion rates (especially when the burning is stable), the
matter reaches high enough temperatures for helium ignition before the
hydrogen has been exhausted. The helium burning then occurs in a
hydrogen rich environment (Lamb \& Lamb 1978; Taam \& Picklum 1978,
1979; FHM; Taam 1982).

For normal or sub-solar metallicities and accretion rates in the
regime $10^{-10} M_\odot {\rm yr^{-1}}\lesssim\dot{M}\lesssim 10^{-8}
M_\odot {\rm yr^{-1}}$, the helium ignition is thermally unstable and
results in a Type I X-ray burst. At higher accretion rates, the helium
burns at a temperature ($T\gtrsim 5\times 10^8\ {\rm K}$) where the
burning is {\it thermally stable} because the nuclear energy
generation rate is less temperature sensitive than the radiative
cooling (FHM). This allows a rough estimate of the critical accretion
rate $\dot{m_{st}}$ where the burning becomes stable (Bildsten 1998b),
\begin{equation}\label{eq:mdotsted}
\dot m_{st}\approx 1.3 \times 10^5 {\rm g \ cm^{-2} \ s^{-1}}\left(M\over 1.4
M_\odot\right)^{1/2}
\left(10 \ {\rm km}\over R\right), 
\end{equation}
in which we take a nominal value $\kappa=0.4\kappa_{es}$, where
$\kappa_{es}=\sigma_{\rm Th}(1+X)/2m_p$ is the Thomson scattering
opacity. Bildsten (1998b) recently reviewed the accretion rate regimes
originally introduced by FHM. The high local accretion rates of the
X-ray pulsars are the standard argument as to why they don't burst
(Joss \& Li 1980, Bildsten \& Brown 1997).

The simple calculation which gives equation (\ref{eq:mdotsted}) is not
accurate enough to say whether $\dot{m}_{st}$ is above or below the
Eddington rate. FHM estimated $\dot M_{st}$ (they called it $\dot
M_{cri}$ in their Table 1) for 3 different neutron stars (here
$\dot{M}=4\pi R^2\dot{m}$).  They found $\dot M_{st}/(10^{-9} M_\odot
\ {\rm yr^{-1}})=15, \ 13, \ {\rm and} \ 12$ for $M=0.476, 1 $ and
$1.41M_\odot$.  Equation (\ref{eq:mdotsted}) agrees to within 20--30\%,
giving $\dot M_{st}/(10^{-9} M_\odot \ {\rm yr^{-1}})=13, \ 16, \ {\rm
and} \ 17$ for the same three masses and radii.  To accurately
determine if the burning is thermally stable requires solving the
time-dependent equations numerically. Ayasli \& Joss (1982) and Taam,
Woosley \& Lamb (1996) performed time-dependent calculations at
accretion rates comparable to the Eddington limit and found that the
burning is thermally unstable when $\dot M < \dot M_{\rm Edd}$.
Ayasli \& Joss (1982) also carried out some calculations with
super-Eddington accretion rates (by a factor of a few) and found
thermal stability.
%{\bf A recent more detailed study by Rembges et al. (1999) finds 
%$\dot M_{st} \approx 0.7 \dot M_{\rm Edd}$
%in agreement with the calculations here.
However, there has not been a thorough survey of the dependence
of the critical accretion rate $\dot m_{st}$ on $M$, $R$ and
composition.

Our focus in this paper is calculating the nucleosynthesis when the
burning is thermally stable. In this regime, the helium is consumed
over a time interval of less than 20 minutes and the resulting
enhancement of the number of CNO seed nuclei together with
breakout from the CNO cycle leads to a rapid consumption of hydrogen
via the rp process.

%--------------------------- SECTION FOUR -------------------------------

\section{Steady-State Burning Models: Helium Ignition as the Hydrogen
Burning Trigger} 
\label{burn_ign}

Steady-state burning of hydrogen and helium takes place in three
phases: hydrogen burning via the hot CNO cycle at low densities (\S
\ref{Burn}), mixed hydrogen and helium burning via the
3$\alpha$ reaction and the $\alpha$p and rp processes when helium
ignites (\S\S \ref{burn_ign} and \ref{burn_hhe}), and pure helium
burning via $\alpha$ captures after the exhaustion of hydrogen (\S
\ref{burn_he}). For all accretion rates where the burning is in
steady-state, consumption of hydrogen by the hot CNO cycle is unimportant
before the helium ignites. It is helium ignition which acts as the
trigger for hydrogen burning. To understand this, we write the depth
at which hydrogen would be consumed by the hot CNO cycle as $y_{\rm
CNO}=\dot m E_{\rm CNO}/\epsilon_{\rm CNO}$, or, using equation
(\ref{eq:cno_burn}),
\begin{equation} \label{eq:yh}
y_{\rm CNO} = 5.74 \times 10^9\ {\rm g\ cm{^{-2}}} \ \dot m_5
\left(\frac{Z_{{\rm CNO},\odot}}{Z_{\rm CNO}}\right).
\end{equation}
For the high accretion rates of interest here, $y_{\rm CNO}$ is always
much greater than the column depth at which helium ignites, $y_{burn}$
(see the estimate of $y_{burn}$ in eq. [\ref{eq:yburn}] below).  Thus
the helium ignites before significant amounts of hydrogen are burned
by the hot CNO cycle\footnote{This does not change even if we
consider breakouts from the hot CNO cycle. We show later in
\S \ref{sec:break} that when breakout occurs at high accretion rates,
the limited number of available seed nuclei and the short time
remaining before the onset of the 3$\alpha$ reaction limits the
hydrogen consumption via breakout reactions prior to helium ignition
to only 10\% of the available fuel.}. Rapid consumption of hydrogen by
the rp process then follows.

\begin{figure}
\epsscale{0.6}
\plotone{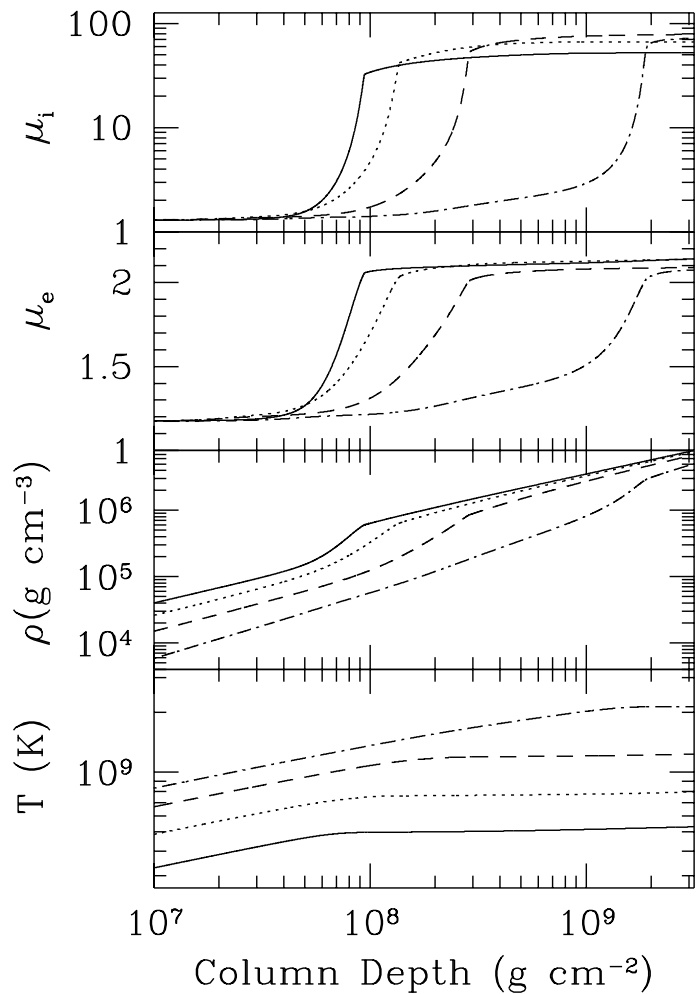}
\caption{
The structure of a rapidly accreting $M=1.4 M_\odot$, $R=10\ {\rm km}$
neutron star atmosphere which is burning its fuel in steady-state. The
plots show the ion mean molecular weight ($\mu_i$), electron mean
molecular weight ($\mu_e$), density and temperature as a function of
the column depth into the star. The solid, dotted, dashed and
dot-dashed lines are for $\dot m/ \dot m_{\rm Edd}=1, 5, 20$ and 50,
respectively.
\label{fig:profiles}}
\end{figure}

\begin{figure}
\plotone{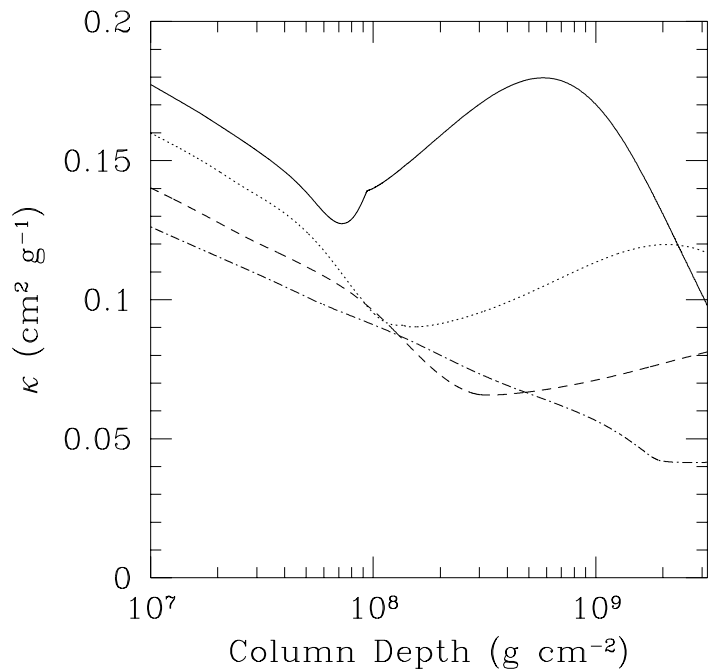}
\caption{
The total opacity $\kappa$ as a function of column depth for the
models shown in Figure~\protect\ref{fig:profiles} and described in
\S~4. The solid, dotted, dashed and dot-dashed lines are for $\dot m/\dot
m_{Edd}=1,5,20$ and $50$ respectively. Above the burning layer, the
opacity is mainly electron scattering. After the main hydrogen burning
occurs, free-free absorption dominates the opacity until conduction by
the degenerate electrons becomes the main heat transport
mechanism. This gives a ``bump'' in the opacity, particularly
prominent at low accretion rates.
\label{fig:opac}}
\end{figure}

We have calculated steady-state models for local accretion rates
between $\dot m/ \dot m_{\rm Edd}=0.7$ and 60. The temperature,
density and composition as a function of column depth are shown in
Figure~\ref{fig:profiles} for four accretion rates, $\dot m/
\dot m_{\rm Edd}=1, 5, 20, $ and 50, which we will discuss in
detail. The steep rise in the mean molecular weight indicates the
location of most of the hydrogen burning. Figure~\ref{fig:opac} shows
the opacity $\kappa$ for these models. The opacity ``bump'',
particularly prominent at low accretion rates, is due to free-free
absorption. This dominates electron scattering after the heavy
elements are made and until the opacity is set by electron conduction
(see Appendix for a full discussion).  Figure~\ref{fig:trho} shows the
temperature-density profiles. The region to the left of the dashed
line is where the temperature is high enough so that the positron
fraction exceeds 10\%.  We have safely stayed in the regime where we
can neglect positrons, confining our calculations to $\dot m/
\dot m_{\rm Edd}<60$.  We show later that this limit coincides with
the onset of nuclear statistical equilibrium, where the outcome of the
nucleosynthesis is simply determined by thermodynamics and does not
need to be calculated in detail.
 
Figure~\ref{fig:abut} shows the abundances of hydrogen and helium as a
function of time and column depth. It is clear that nearly all the
hydrogen burning occurs where the helium ignites. We make a simple
estimate of this location by matching the helium lifetime to the
$3\alpha$ reaction to the time it takes to cross a scale height. Using
the energy generation rate for helium burning via the 3$\alpha$
reaction ($\epsilon_{3\alpha}$), we write $\dot m Y/y \approx
\epsilon_{3\alpha}/E_{3\alpha}$ (Taam 1981, Fushiki \& Lamb 1987,
Bildsten 1995) where $E_{3\alpha}=5.84\times 10^{17} \ {\rm ergs \
g^{-1}}$ is the energy release from $3\alpha \rightarrow
^{12}$C. Using the temperature-density relation for an atmosphere with
pure Thomson scattering and a constant flux $F=10^{23} \ {\rm erg \
cm^{-2}\ s^{-1}} E_{18} \dot m_5$, we find that the helium ignition
temperature is given by the solution to
\begin{equation}\label{eq:trans}
\dot m_5^4=68.2\left[{{(Yg_{14} \mu)^2 T_8^7} \over {(1+X)^3E_{18}^3}}\right]
\exp\left({-44\over T_8}\right)
\end{equation}
(Bildsten 1998b), where we write the energy released per gram of
accreted material in units of $10^{18} \ {\rm erg \ g^{-1}}$ as
$E_{18}$ and the accretion rate in units of $10^5 {\rm g \ cm^{-2} \
s^{-1}}$ as $\dot m_5$. Solving this transcendental equation for the
temperature gives the pressure at which the helium burns and, most
importantly, how these quantities depend on $\dot m$, $M$, $R$ and the
abundances. In order to simplify the transcendental, we expand the
exponential about the temperature of $T_8=6.285$ (in the middle of the
range we are interested in) so that $\exp(-44/T_8)\approx 9.11\times
10^{-4}(T_8/6.285)^{7}$.  Some accuracy is compromised because of this
approximation, but not much. We thus find
\begin{equation}\label{eq:tburn}
T_{\rm burn}^{\rm est}=3.06
\times 10^8 {\rm K} \ \dot m_5^{2/7} {{(E_{18}+XE_{18})^{3/14}} \over 
(Yg_{14}\mu)^{1/7}},
\end{equation} 
and
\begin{equation}\label{eq:yburn}
y_{\rm burn}^{\rm est}={{3.36\times 10^7 \ {\rm g\ cm^{-2}} \dot
m_5^{1/7}}\over {(E_{18}+XE_{18})^{1/7}(Y g_{14} \mu)^{4/7}}}.
\end{equation}
as the temperature and column density at the helium burning location.

These estimates compare well with our numerical results (we take
$E_{18}=5.3$, $g_{14}=1.86$, $\mu=0.61$, $X=0.71$, and
$Y=0.28$). Equation (\ref{eq:yburn}) accurately gives the location
where $10-20$\% of the helium is burned. The temperature estimate is
consistently 20\% too high (for the same location).  We have also
checked the dependence of the burning conditions on gravity for a
fixed $\dot m=7.5\times 10^4 \ {\rm g \ cm^{-2} \ s^{-1}}$ model. We
find that the burning takes place at higher temperatures, lower
densities and higher column depths for a lower gravity, confirming the
gravity dependence of equations (\ref{eq:tburn}) and (\ref{eq:yburn}).
Thus our simple estimate accurately reproduces the scalings seen in
the numerical simulations.

% ------------------------ SECTION FIVE ----------------------------------
\section{Nucleosynthesis from Mixed Hydrogen and Helium Burning}
\label{burn_hhe}

When the helium ignites, mixed hydrogen and helium burning occurs via
the 3$\alpha$ reaction, the $\alpha$p process and the rp process. The
ignition of $3\alpha$-burning and subsequent carbon production in a
hydrogen rich environment provides additional seeds which accelerate
the burning of hydrogen. The $\alpha$p process is a series of
($\alpha$,p) reactions and proton captures starting with the
$^{14}$O($\alpha$,p)$^{17}$F reaction. This is a pure helium burning
process since for each proton released in an ($\alpha$,p) reaction a
proton is captured in a subsequent (p,$\gamma$) reaction (Wallace and
Woosley 1981).  Hydrogen must be present for the $\alpha$p process to
play a role, since proton capture reactions provide the link between
the $^{12}$C produced by the 3$\alpha$ reaction, and the $^{14}$O at
the beginning of the $\alpha$p process.  Most of the hydrogen is
burned by the rp process (rapid proton capture process), a series of
fast proton captures and slow $\beta$-decays close to the proton drip
line (Wallace and Woosley 1981).  We show later that the typical mass
of nuclei at the endpoint of the rp process ($A_{\rm rp}$) is around
60--100 and is determined both by the exhaustion of hydrogen and the
prevalence of the $\alpha$p process.

Before describing our detailed results, we start with a simple picture
of the mixed hydrogen and helium burning. Imagine that helium burns
only via the 3$\alpha$ reaction and breakout from the hot CNO cycle,
for example via the $^{15}$O($\alpha$,$\gamma$)$^{19}$Ne
reaction. Then, $^{19}$Ne is the seed nucleus for hydrogen burning via
the rp process. Neglecting the small amount of hydrogen burned before
helium ignition, for solar abundance there are 38 protons per
$^{19}$Ne seed nucleus.  If there are no catalytic loops, the ensuing
rp process would produce nuclei with a typical mass $A_{\rm rp}\approx
57$ after all the hydrogen is burned.  We generalize this simple
estimate to the case where helium is burning via the $\alpha$p
process, by assuming that only one endpoint nucleus is produced from
the $\alpha$p process (mass number $A_{\alpha p}$), which again serves
as the seed nucleus for the rp process. The typical mass $A_{\rm rp}$
of the endpoint of the rp process then depends on the mass fraction of
helium burned into seed nuclei for the rp process ($Y_{\rm burn}$) and
the endpoint of the helium burning ($A_{\alpha p}$),
\begin{equation} \label{eq:arp}
A_{\rm rp} \approx A_{\alpha p} \left( 1 + \frac{X}{Y_{\rm burn}}
\right)
\end{equation}
(Schatz et al. 1998). 
Thus the strong temperature sensitivity of the
$(\alpha$,p) reactions in the $\alpha$p process is imprinted on the
rp process ashes. 
In this section, we first describe the physics that
determines $A_{\alpha p}$ and show that the temperature at the depth
of helium ignition is the most important parameter. We then discuss
the rp process and the resulting distribution of elements, and show
that equation (\ref{eq:arp}) is a good estimate of our numerical
results.

\subsection{CNO Breakout and the $\alpha$p Process }
\label{sec:break}

Crucial for the rp process is the extent to which $\alpha$-induced
reactions other than the 3$\alpha$ reaction can take place. The major
limiting factor is the Coulomb barrier for the $\alpha$ particles,
which increases rapidly towards heavier nuclei. In Figure~\ref{tr},
the dotted line shows the temperatures and densities where the
$^{15}$O($\alpha$,$\gamma$)$^{19}$Ne rate equals the $^{15}$O
$\beta^+$ decay rate. To the right of this line, breakout from the hot
CNO cycle can occur. The dashed lines show where the ($\alpha$,p)
reaction rate on $^{14}$O, $^{18}$Ne, $^{22}$Mg, $^{26}$Si, $^{30}$S
and $^{34}$Ar equals the rate of destruction (by $\beta^+$ decay and
proton capture) of each of these isotopes. The $\alpha p$ process will
proceed via a particular $(\alpha$,p) reaction if the temperature and
density of a fluid element reach the corresponding threshold (dashed
line) during helium burning. Otherwise, the $\alpha p$ process will
end and the reaction flow towards heavier nuclei will continue via the
rp process.

For nuclei above $^{18}$Ne proton captures can in principle compete
with $\alpha$-induced reactions as the proton capture rates are
typically much faster than the ($\alpha$,p) reaction rates. However,
at the temperatures where the helium burning occurs, the ($\gamma$,p)
photodisintegration reactions are in balance with the (p,$\gamma$)
reactions, which strongly hampers effective proton captures. This does
not mean, as it is often concluded, that proton captures are always
negligible. In fact, the deviations of the thresholds shown in
Figure~\ref{tr} from the smooth behavior at low densities are due to
proton captures. For example, let's consider $^{22}$Mg. In this case,
the nuclei $^{22}$Mg and $^{23}$Al are in (p,$\gamma$)-($\gamma$,p)
equilibrium determined by the Saha equation. However, the
$^{23}$Al(p,$\gamma$)$^{24}$Si reaction operating on the low $^{23}$Al
equilibrium abundance establishes a net reaction flow from $^{22}$Mg
to $^{24}$Si.  This is similar to the 2p capture reactions discussed
in G\"orres et al. (1995) and Schatz et al. (1998). This type of
proton capture is a weak process but goes with $\rho^2$, and therefore
there is a critical density above which the ($\alpha$,p) reaction
competes with proton captures rather than with the density independent
$\beta$ decay. For $^{22}$Mg this happens at densities above $10^4\
{\rm g\ cm^{-3}}$. As can be seen in Figure~\ref{tr}, the threshold
for the ($\alpha$,p) reaction behaves very differently in that regime
because of the very different density and temperature dependence of
the competing reaction. For $^{22}$Mg, $^{26}$Si, $^{30}$S, and
$^{34}$Ar, the critical density is such that proton captures play an
important role in determining the $\alpha$p process path at these
isotopes.  For example, proton captures move the border for the
$^{22}$Mg($\alpha$,p)$^{25}$Al reaction to higher temperatures, which
is why the $\alpha$p process needs accretion rates above 20$\
\dot{m}_{\rm Edd}$ to proceed beyond $^{22}$Mg. Proton captures also
lead to an inversion of the thresholds for the
$^{26}$Si($\alpha$,p)$^{29}$P and the $^{30}$S($\alpha$,p)$^{33}$Cl
reactions for densities above $10^5\ {\rm g\ cm^{-3}}$.  On the other
hand, in the cases of $^{14}$O and $^{18}$Ne the next isotone is
proton unbound and the necessary densities are much higher, roughly
$10^9{\rm g\ cm^{-3}}$ (G\"orres et al. 1995).

The solid lines in Figure~\ref{tr} are the evolutionary tracks for
fluid elements in the temperature-density plane at various accretion
rates. The fat line segment indicates where hydrogen burning occurs
(from 90\% to 10\% of the initial hydrogen abundance) and clearly
shows that the temperature-density track for all accretion rates
crosses the threshold for the $^{15}$O($\alpha$,$\gamma$)$^{19}$Ne
reaction before exhausting the hydrogen. Thus at all accretion rates,
there is a breakout of the hot CNO cycle. There are uncertainties
in the $^{15}$O($\alpha$,$\gamma$)$^{19}$Ne reaction rate by factors
of 5--10 (Mao et al. 1995, 1996, Wiescher, G{\"o}rres \& Schatz
1999). However, the strong temperature dependence of the reaction rate
means that changes in the reaction rate within the uncertainties have
hardly any impact on the reaction flow and the breakout of the hot CNO
cycle. To confirm this, we have performed a test calculation with
$\dot m/ \dot m_{\rm Edd}=0.7$ and the
$^{15}$O($\alpha$,$\gamma$)$^{19}$Ne reaction rate reduced by a factor
of ten. The only difference we find in this case is that the column
depth of the burning zone is deeper by 8\% resulting in a slightly
longer CNO burning phase before the breakout. As we discuss in \S
\ref{burn_he}, this slightly increases the amount of helium that
survives hydrogen burning. {\it Therefore we conclude that steady
state hydrogen burning must always take place via the rp process.}

The time-integrated reaction flows for accretion rates of $\dot m/
\dot m_{\rm Edd}=1$, 5, 20, and 50 are shown in Figure~\ref{flow}.  At
accretion rates below $\dot m_{\rm Edd}$, the only important
$\alpha$-induced reaction (apart from the 3$\alpha$ process) is
$^{15}$O($\alpha$,$\gamma$)$^{19}$Ne, which triggers the breakout of
the hot CNO cycle into the rp process starting at $^{19}$Ne.  The
$^{14}$O($\alpha$,p)$^{17}$F reaction requires somewhat higher
temperatures and starts to occur at $\dot m/ \dot m_{\rm Edd}\approx
1$. This reaction does not lead to a breakout of the hot CNO cycle but
instead opens up another loop that proceeds via
$^{17}$F(p,$\gamma$)$^{18}$Ne($\beta^+$)$^{18}$F(p,$\alpha$)$^{15}$O. This
loop just provides another path from $^{14}$O to $^{15}$O having the
net effect of a $\beta$ decay and a proton capture.  The major
hydrogen burning at $\dot m = \dot m_{\rm Edd}$ still proceeds via the
$^{15}$O($\alpha$,$\gamma$)$^{19}$Ne reaction and the subsequent
synthesis of heavier elements via the rp process. The picture changes
at higher accretion rates, however, where the
$^{18}$Ne($\alpha$,p)$^{21}$Na reaction becomes the dominant breakout
of the hot CNO cycle. The reaction flow from the CNO region towards
heavier elements is then dominated by the $\alpha$p process starting
with the $^{14}$O($\alpha$,p)$^{17}$F reaction. As can be seen in
Figure~\ref{tr}, as $\dot m$ increases the ignition temperature gets
higher and more ($\alpha$,p) reactions become possible.  This is
confirmed in Figure~\ref{mrange}, which shows the mass number of the
last isotope made in the $\alpha$p process $A_{\alpha \rm p}$ as a
function of $\dot m$. This mass increases with accretion rate and for
$\dot m/ \dot m_{\rm Edd}=50$ the $\alpha$p process does not end until
after $^{41}$Sc. The rapid increase in $A_{\alpha \rm p}$ at $\dot
m>15\ \dot m_{\rm Edd}$ reflects the small temperature differences
between the thresholds for the various ($\alpha$,p) reactions at
$\rho>10^5\ {\rm g\ cm^{-3}}$ (Figure~\ref{tr}). The inversion of the
$^{26}$Si($\alpha$,p)$^{29}$P and the $^{30}$S($\alpha$,p)$^{33}$Cl
reaction thresholds is reflected in the jump in $A_{\alpha \rm p}$
from 26 to 34 just above 20$\ \dot m_{\rm Edd}$. As mentioned earlier,
these effects are a consequence of the competition between proton
captures and ($\alpha$,p) reactions.

Figure~\ref{abut2} shows the abundances of hydrogen, helium,
carbon, oxygen, nickel and the last major isotope produced in the rp
process for several accretion rates as a function of time. Also shown
is the abundance of one of the isotopes that dominates the composition
at the end of our calculation. At $\dot m=\dot m_{\rm Edd}$, the CNO
cycle opens only at $^{15}$O, which causes the slight drop in the
$^{15}$O abundance and the build up of $^{56}$Ni after about 200~s
($^{56}$Ni is the first major waiting point in the rp process
path). The destruction of $^{15}$O is then delayed by the conversion
of $^{14}$O into $^{15}$O via the hot CNO cycle.  At higher accretion
rates the opening of the CNO cycle is reflected in the rapid depletion
of $^{14}$O and $^{15}$O with the simultaneous production of heavier
nuclei like $^{56}$Ni. There is a brief period of time between the
breakout of the CNO cycle and the onset of the 3$\alpha$ reactions
supplying additional CNO nuclei. This is best seen for $\dot m=50\
\dot{m}_{\rm Edd}$, where the first peak in the $^{56}$Ni abundance
comes from processing of the initial CNO abundances, while the second
peak at 20~s indicates the onset of the 3$\alpha$ reaction supplying
more seed nuclei. The time between the breakout from the CNO cycle and
helium ignition ranges from roughly 260~s at $\dot m=\dot m_{\rm Edd}$
to 20~s at $\dot m = 50\ \dot m_{\rm Edd}$. However, the amount of
hydrogen burned in this period via the rp process is only about
$8-10$\% of the initial hydrogen abundance because of the relatively
long timescales for rp processing. Therefore, helium ignition always
takes place in a hydrogen-rich environment.

The build up of the last isotope in the rp process path (dot-dashed
line in Figure \ref{abut2}) coincides with the depletion of
hydrogen. The peak abundances reached for these isotopes are quite low
as they are determined by the competition between production by the rp
process and rapid destruction via fast $\beta$ decays.  Once the
hydrogen is gone, the isotopes along the rp process path quickly decay
into more stable isobars via a series of $\beta$ decays. The resulting
build up of the most abundant nuclei at the end of our calculation is
shown as dotted lines in Figure \ref{abut2}.  Some of these nuclei are
still unstable and will $\beta$ decay on longer timescales at greater
depths.

\subsection{The rp Process, Hydrogen Exhaustion and Residual Helium} 
\label{sec:rp}

The nuclei produced by helium burning are the seeds for burning the
hydrogen via the rp process. As discussed above, the burning
conditions are set by the point of helium ignition. For each accretion
rate, the thick line segment in Figure~\ref{tr} shows the region of
the temperature-density plane where the hydrogen burns. The density is
roughly the same for all accretion rates and increases from about
$10^5\ {\rm g\ cm^{-3}}$ to $10^6\ {\rm g\ cm^{-3}}$ during the
burning. On the other hand, the burning temperature depends strongly
on the accretion rate and increases from $5\times 10^8$~K at $\dot
m=\medd$ to $15$--$20\times 10^8$~K at $\dot m=50\medd$. The timescale
for hydrogen burning is given in Table~\ref{endpoints} and varies
between 100 and 500 s.

Figure~\ref{mrange} shows the mass number of the rp process endpoint
nucleus, $A_{\rm rp}$, as a function of $\dot m$. The corresponding
reaction flows for some selected cases are in
Figure~\ref{flow}. Clearly, $A_{\rm rp}$ increases with $\dot m$ and,
for $\dot m> 10\ \dot m_{\rm Edd}$, the rp process reaches the end of
our reaction network near $A=100$. As discussed in \S \ref{summary} we
expect that for accretion rates above $10 \ \dot m_{\rm Edd}$ (but
less than $50 \ \dot m_{\rm Edd}$, see discussion below) significant
amounts of nuclei heavier than $A=100$ are produced.

The rp process reaction path in Figure~\ref{flow} is characterized by
a sequence of proton captures and $\beta$-decays and proceeds above
$^{56}$Ni along the $N=Z$ line as discussed in Schatz et al. (1998).
However, we find that two-proton-capture reactions on $^{68}$Se and
$^{72}$Kr do not play a role in steady state burning owing to the somewhat
lower densities and temperatures compared to the X-ray burst peak conditions
discussed in Schatz et al. (1998).
Some branchings occur when proton captures are slowed down by
photodisintegration or, at lower accretion rates and temperatures,
because of small reaction rates.  At accretion rates above $45\
\dot{m}_{\rm Edd}$ the temperatures get so high ($T_8 > 20$) that
photodisintegration starts to severely hamper the rp process. At $\dot
m = 50 \dot m_{\rm Edd}$, photodisintegration of $^{84}$Mo inhibits
further proton captures on $^{83}$Nb. As a consequence, the
$^{83}$Nb(p,$\alpha$)$^{80}$Zr reaction becomes the dominant
destruction mechanism of $^{83}$Nb and the rp process ends in the
Zr-Nb cycle.  This causes the drop in $A_{\rm rp}$ in
Figure~\ref{mrange} at high $\dot m$'s.  As discussed in Schatz et
al. (1998) the Zr-Nb cycle occurs at high temperatures because of the
very low $\alpha$ binding energy of $^{84}$Mo. This low $\alpha$
binding energy is predicted by the FRDM (1992) mass model (M\"oller et
al. 1995) used in our study. An experimental confirmation of this
would be highly desirable. At accretion rates beyond $\dot m=50 \dot
m_{\rm Edd}$, the temperatures get so high that photodisintegration
drives the material into nuclear statistical equilibrium. The nuclei
produced in the rp process are then quickly converted into iron peak
isotopes, mostly $^{56}$Ni. {\it It is only in this very high
accretion rate regime ($\dot m>50\ \dot m_{\rm Edd}$) that we find a
pure nickel solution that will later decay to make the ashes nearly
pure iron. }

What determines the endpoint of the rp process for accretion rates
$\dot m<50\ \dot m_{\rm Edd}$? The rp process ends either when all the
hydrogen is consumed or when the proton capture rates become too slow
because of the increasing Coulomb barrier.  The effect of the Coulomb
barrier can be estimated from the formula given in Woosley and Weaver
(1984)
\begin{equation} \label{eq:zmax}
Z_{\rm max}^{2/3} \approx \ln \left[ \frac{1.5 \times 10^{11} \rho X \tau}
 {T_9^{2/3}} \right] \frac{T_9^{1/3}}{4.25-1.33 T_9^{1/3}}, 
\end{equation}
where $Z_{\rm max}$ is the maximum charge number that can be
synthesized in a time $\tau$ at a density $\rho$, a temperature $T_9$
(in 10$^9$~K) and a proton abundance $X$. We find that, for typical
conditions during hydrogen burning and using our timescale for
hydrogen burning from Table 1, the $Z_{\rm max}$ is always higher than
the heaviest nucleus produced in our calculations: for $\dot m / \dot
m_{\rm Edd}=1$, 5, 20, 50 we find $Z_{\rm max}=38$, 45, 61, and
92. {\it Therefore, the Coulomb barrier does not limit the rp process
in steady state burning at high accretion rates.} This is also
confirmed from the reaction flows in Figure~\ref{flow}. The flows are
always close to the proton drip line all the way up to the heaviest
nucleus synthesized, indicating that it is the $\beta$-decays and not
the proton captures that set the timescale. The deviation of the rp
process path from the proton drip line by 1 or 2 mass units is in most
cases due to photodisintegration of the weakly-bound proton-rich
nuclei.

We find that hydrogen is completely consumed by the rp process
near the depth where helium ignites. No hydrogen reaches deeper
regions of the atmosphere, and electron capture on hydrogen does not
occur when the burning is in steady-state, as earlier speculated by
Taam et al. (1996). The only exception is at very high accretion
rates, typically beyond $50\ \dot{m}_{\rm Edd}$, where
photodisintegration hampers efficient hydrogen burning via the rp
process beyond $^{56}$Ni. Indeed, Figure~\ref{flow} shows that, at
$\dot m=50\medd$, there is a very weak reaction flow via hydrogen
electron capture. But even at $\dot m=60\medd$ the hydrogen mass
fraction that is left after the rp process burning is 0.4\%.  
These results contradict the speculations of Taam et al. (1996) about the
possible stabilizing effects of hydrogen electron captures at large
depths. A likely explanation for this is that the network of Taam et al.
did not include nuclei with $A>56$. However, from the simple arguments
giving equation (13), it is clear that the rp process during steady-state
burning always leads to production of nuclei beyond iron. Thus the omission
of nuclei beyond $A=56$ would result in under-consumption of hydrogen, and
lead one to the incorrect conclusion that hydrogen survives the nuclear burning.

Since it is hydrogen consumption that limits the rp process, the
endpoint is roughly given by equation (\ref{eq:arp}). The important
parameters are the endpoint of the $\alpha$p process and the amount of
helium burned into seed nuclei during hydrogen burning ($Y_{\rm
burn}$). We have shown that $A_{\alpha \rm p}$ can in principle be
obtained from Figure~\ref{tr}, but $Y_{\rm burn}$ is more difficult to
estimate, since some mass fraction of helium ($Y_r$) survives the
hydrogen burning phase (see Table~\ref{endpoints}). This helium is not
available for the production of rp process seed nuclei and burns later
in the pure helium burning phase discussed in \S
\ref{burn_he}.

Table~\ref{endpoints} shows that the amount of helium surviving
hydrogen burning is 23\% of the initial helium abundance at $\dot
m=\dot m_{\rm Edd}$. Helium survives the hydrogen burning because
the timescale to burn helium via the 3$\alpha$-reaction,
$\tau_\alpha$, is longer than the timescale for reaching the last
isotope in the rp process, $\tau_{\rm rp}$, which is the sum of the
lifetimes of all the nuclei along the calculated reaction paths.
Figure~\ref{tau} shows $\tau_\alpha$ and $\tau_{\rm rp}$ as a function
of mass number for $\dot m/\dot m_{\rm Edd}=1$ and 20. When
calculating $\tau_{\rm rp}$, we neglect the influence of
photodisintegration and of the change in hydrogen abundance during the
burning. This is a reasonable approximation since the rp process
timescale is mostly set by $\beta^+$ decays. Only the proton captures
at $^{52}$Fe and $^{56}$Ni contribute to the rp process timescale at
low accretion rates. At low $\dot m$, the timescale for helium burning
is indeed significantly longer than the timescale for the rp process
to reach its endpoint and burn all the hydrogen.  At higher accretion
rates, the two timescales become comparable, and the amount of helium
remaining unburned after hydrogen exhaustion is less.

\subsection{The Final Abundance Distribution}

The final abundance distribution of the rp process ashes from steady
state burning is shown in Figure~\ref{abu} for $\dot m/ \dot m_{\rm
Edd}=1$, 5, 20, 50, and 60. As discussed above, the higher the
accretion rate, the heavier the nuclei produced until, at $\dot
m=50\medd$, the rp process gets stuck in the Zr-Nb cycle at $A=80$. At
this accretion rate, nuclear statistical equilibrium starts to play a
role during the final burning stage, and drives some nuclei back into
$^{56}$Ni. These two effects lead to the ``double peak'' structure of
the abundance pattern with maxima around $A=56$ and $A=80$. At $\dot
m=60\medd$, nuclear statistical equilibrium dominates and all nuclei
are converted into iron peak nuclei, mainly $^{56}$Ni.

The abundance distribution for $\dot m< 50\ \dot m_{\rm Edd}$ is
determined by the nuclei with the slowest reaction rates in the rp
process path, the so-called ``waiting points'' (Wallace and Woosley
1981). Some fraction of the material is locked into these waiting
points until the burning is over and this leads to the production of a
wide range of isotopes. The important waiting points for the
conditions discussed here have been identified in Schatz et al. (1998)
and are the even-even $N=Z$ nuclei $^{56}$Ni, $^{64}$Ge, $^{68}$Se,
$^{72}$Kr, $^{76}$Sr and $^{80}$Zr. For these nuclei, proton captures
are inefficient since photodisintegration or proton decay can remove a
captured proton quickly and $\beta$-decay rates are relatively low.
This explains the local maxima in the abundance distributions (for
$\dot m/\dot m_{\rm Edd}=1-20$) at the mass numbers of these waiting
point isotopes.  Additional peaks occur at the nuclei that are
constructed by $\alpha$ capture on $^{12}$C as discussed in
\S \ref{burn_he}.  At $\dot m=\medd$, an additional peak at $A=52$ occurs
because of the relatively slow $^{52}$Fe(p,$\gamma$)$^{53}$Co reaction
rate at low temperatures (this can also be seen in Figure~\ref{tau},
where for low accretion rates the first major increase of the rp
process timescale occurs at $A=52$). The flat and structureless
abundance distribution above $A=72$ in the $\dot m=5\medd$ case is
caused by the fact that the rp process barely leaks beyond the major
waiting points $^{68}$Se and $^{72}$Kr. As can be seen in
Figure~\ref{flow}, the drastically reduced processing timescale beyond
$^{72}$Kr (see Schatz et al. 1998) distributes the reaction flow over
a wide range of nuclei as hydrogen is rapidly depleted.

\subsection{Summary of the Mixed Burning and Energy Generation}\label{summary}

In Figure~\ref{mrange}, we show the mass number of the last isotope
reached in the $\alpha$p process ($A_{\alpha p}$, open circles) and in
the rp process ($A_{\rm rp}$, open squares) as a function of $\dot
m$. Also shown is the average mass number ($<A>$, filled circles) and
the average charge number ($<Z>$, filled triangles) of the final
composition.  We now test equation (\ref{eq:arp}) for accretion rates
$\dot m<50\medd$ by assuming $Y_{\rm burn}=Y-Y_r$ and by taking $Y_r$
from Table~\ref{endpoints}. The resulting estimated endpoints of the
rp process are shown in Figure~\ref{mrange} as open triangles.  As
expected the estimated endpoint is a few mass numbers too high, since
some hydrogen is burned during the beginning of the burning phase when
temperatures are lower and the $\alpha$p process can not yet reach its
final endpoint.

Thus the estimate of equation (\ref{eq:arp}) gives an upper limit to
the rp process endpoint in the accretion rate regime between 15 and 45
$\dot m_{\rm Edd}$ where the reaction flow reaches the end of our
network. Figure~\ref{mrange} shows that the heaviest nuclei that can
be produced by the rp process in steady state burning are estimated to
be in the $A=150$ region. At these mass numbers the proton drip line
is around $Z=70$ and applying equation (\ref{eq:zmax}) we find that
the Coulomb barrier should not inhibit the rp process even for these
heavy nuclei. However, the endpoint of the rp process might be
considerably below this upper limit because of loops in the reaction
path that might occur above $^{100}$Sn as the reaction path enters a
region of nuclei with very low $\alpha$ binding energies.

{\it To summarize, the synthesis of much heavier isotopes at higher
accretion rates is not because of faster proton capture rates, but due
to enhanced ($\alpha$,p)-reactions that lead to a longer $\alpha$p
process and therefore to heavier seed nuclei and a larger fuel to seed
ratio for the rp process.} The final abundance distributions as a
function of mass number are shown in Figure~\ref{abu} for various
accretion rates. {\it Generally, not just a single nucleus but a whole
range of isotopes is produced by steady state burning}. Even for our
lowest accretion rate of $\mdot=0.7\medd$ significant amounts of
nuclei heavier than $^{56}$Ni are synthesized.

The mixed hydrogen and helium burning is responsible for nearly all of
the released nuclear energy. The energy released by the CNO cycle
before helium ignition (\S \ref{burn_ign}) and by the helium burning
after hydrogen consumption (\S \ref{burn_he}) is
negligible. Table~\ref{endpoints} lists the energy produced by nuclear
burning for different accretion rates. Hydrogen and helium are
completely burned in all of our models. Because of the weak dependence
of nuclear binding energy with mass number in the range of the rp
process endpoints reached in this study, the released energy depends
only weakly on accretion rate and burning conditions. This energy is
predominantly produced by hydrogen burning.  Helium burning does not
contribute more than 0.6~MeV/nucleon (3$\alpha$ only) to
1.2~MeV/nucleon ($\alpha$p process up to $^{37}$K).

\subsection{Helium Burning after Hydrogen Exhaustion} 
\label{burn_he}

It was pointed out previously (in \S \ref{sec:rp}) that the relatively
long timescale for helium burning allows some helium to survive the
hydrogen burning phase. The mass fractions of unburned helium, $Y_r$,
for various accretion rates are listed in Table \ref{endpoints}. This
residual helium burns later in a hydrogen-free environment via the
3$\alpha$ reaction and, depending on the temperature, via a series of
$\alpha$ captures starting at $^{12}$C. Figure \ref{abut2} shows the
build up of $^{12}$C right after the hydrogen is burned, and, at
accretion rates above $\mdot=\medd$, the depletion of $^{12}$C by the
subsequent $^{12}$C($\alpha$,$\gamma$)$^{16}$O reaction. Following the
$^{12}$C($\alpha$,$\gamma$)$^{16}$O reaction, a sequence of $\alpha$
captures occurs that can be identified in Figure~\ref{flow}. This
$\alpha$ capture chain is longer for the higher temperatures at higher
accretion rates, although less helium is then available (see
Table~\ref{endpoints}).  At $\mdot=20\medd$, the $\alpha$ capture
chain reaches $^{44}$Ti. The nuclei produced during the pure helium
burning phase can be identified in the abundance patterns shown in
Figure~\ref{abu} and are $^{12}$C and $^{24}$Mg for $\mdot=\medd$,
$^{12}$C, $^{28}$Si, and $^{32}$S for $\mdot=5\medd$ and $^{40}$Ca and
$^{44}$Ti for $\mdot=20\medd$.

The final mass fraction of $^{12}$C for accretion rates of $\dot m /
\dot m_{\rm Edd}=1$ and 5 is 4.1\% and 0.23\%, and is essentially zero
for higher accretion rates. The final amount of $^{12}$C is much
larger at low accretion rates, because more helium is available after
hydrogen exhaustion (Table~\ref{endpoints}) and at lower temperatures
much less $^{12}$C is destroyed by the
$^{12}$C($\alpha$,$\gamma$)$^{16}$O reaction.

These conclusions depend somewhat on the rates for the CNO breakout
reactions. For low accretion rates, where most $^{12}$C is produced,
this is the $^{15}$O($\alpha$,$\gamma$)$^{19}$Ne reaction. In
principle, a lower breakout rate leads to more hydrogen burning via
the hot CNO cycle, in which helium is produced (in contrast to the rp
process), giving a higher final helium abundance after hydrogen
burning. However, we find that a factor of ten lower
$^{15}$O($\alpha$,$\gamma$)$^{19}$Ne reaction rate increases the final
$^{12}$C mass fraction by just 10\% (a factor of ten higher rate
reduces it by 25\%; the uncertainty in the reaction rate is a factor
of 5--10, see \S \ref{sec:break}). We conclude that the final mass
fraction of $^{12}$C produced in steady state burning will never
exceed 6\%. It thus seems unlikely that enough $^{12}$C is produced to
release additional energy via explosive carbon burning deeper in the
atmosphere as discussed by Brown \& Bildsten (1998).

%-------------------------- CONCLUSIONS ----------------------------------

\section{Conclusions}

We have fully explored the complicated nucleosynthesis from thermally
stable mixed hydrogen/helium burning at high accretion rates on
accreting neutron stars. Our major finding is that the rp process
produces a mixture of very heavy elements, the average mass of which
depends on the local accretion rate.

The important reaction sequences during the mixed hydrogen and helium
burning are the 3$\alpha$ reaction, the $\alpha$p process and the rp
process. For all accretion rates where the nuclear burning is in
steady state, a breakout of the hot CNO cycles into the rp process
takes place. It is the endpoint of the rp process that determines the
energy generation and final composition of the ashes. In contradiction
to Taam et al. (1996), we find that no hydrogen survives the steady
state burning and consequently deep hydrogen burning by electron
capture does not take place (the possibility still remains that there
is substantial residual hydrogen from unstable burning in X-ray
bursts). For $^{12}$C the situation is similar: for accretion rates of
a few times Eddington, some $^{12}$C survives the burning, but never
enough to trigger carbon flashes via exposive burning in deeper layers
of the atmosphere.

The most important nuclear physics input parameters for our
steady state burning calculations are 1) the 3$\alpha$ reaction rate,
which triggers the burning;
2) the breakout reactions from the hot CNO cycles
$^{15}$O($\alpha$,$\gamma$)$^{19}$Ne and $^{18}$Ne($\alpha$,p)$^{21}$Na,
as discussed in \S 5.1; 3) the ($\alpha$,p) and (p,$\gamma$)-
reaction rates on $^{14}$O, $^{22}$Mg, $^{26}$Si, and $^{30}$S together with
the proton separation energies and the proton capture rates
of $^{23}$Al, $^{27}$P, $^{31}$Cl, and
$^{35}$K (these data affect the extent of helium burning and therefore
the endpoint of the rp process); 4) the proton capture rates on the
waiting points $^{52}$Fe and $^{56}$Ni; and 5) proton capture Q-values
and $\beta$-decay half lives of the even $Z$, $N=Z$ and the even $Z$,
$N=Z+1$ nuclei between $^{56}$Ni and $^{100}$Sn as discussed in
Schatz et al. (1998). While the 3$\alpha$ reaction is known with
sufficient accuracy, most of the other data are completely or
partially based on theoretical data, for which extrapolation to the
very neutron deficient nuclei in the rp process is often doubtful
(see Schatz et al. 1998 for a more detailed discussion).
However, while these uncertainties may affect the detailed abundance
pattern for a given local accretion rate, they have no influence on our
general conclusions concerning the nature of steady state burning on
accreting neutron stars (see \S 5.1). 
Nevertheless, more experimental information on the nuclear data mentioned
above would certainly be desirable.

Our calculations of the composition of the nuclear ashes in
steady-state burning show that the ocean and crust of an accreting
neutron star do not consist of pure iron, as assumed in previous
work. Instead, the final composition consists of a wide range of
nuclei. This is characteristic of the rp process in which some
fraction of nuclei is locked at a large number of waiting points with
long lifetime.

Our results will have interesting consequences for studies of the
crust of accreting neutron stars. For example, our discovery of the
large range of nuclei present will directly impact estimates of the
thermal and electrical conductivity of the crust.  To illustrate this
point, we have calculated the ``impurity parameter'' $Q=Y_{\rm max}^{-1}
\sum_j Y_j (Z_j-Z_{\rm max})^2$ for our models, where $Y_j$ are the
nuclear abundances and the subscript max indicates the most abundant
species. We find that $Q\sim 100$ is typical of the mixture from the
ashes of steady state burning. Even at very high accretion rates,
($\mdot\gtrsim 50\medd$) when nuclear statistical equilibrium favors
$^{56}$Ni as the sole product of nucleosynthesis, $Q\approx 1$. Brown
\& Bildsten (1998) showed that for $Q\ge 1$ the thermal and electrical
conductivity is dominated by impurity scattering, which is strongly
composition dependent. In previous work which assumed a crust of pure
iron (before electron captures), impurity scattering was
unimportant. Thus in accreting neutron stars, the thermal and
electrical crust conductivities will be in general much lower than
previously assumed, leading to a different thermal structure and
faster Ohmic diffusion of magnetic fields in the crust.

\acknowledgements 

We thank Ed Brown and Felix Rembges for many discussions during this
work. We are especially grateful to F.-K. Thielemann and T. Rauscher
for providing the reaction network solver and many of the reaction
rates. This research was supported by NASA via grants NAG 5-2819 and
NAGW-4517. L. B. was supported by the Hellman Family Faculty Fund
Award (UCB) and the Alfred P. Sloan Foundation.

%----------------------------- APPENDIX ---------------------------------

\appendix
\section{Radiative and Conductive Opacities}

Electron scattering, free-free absorption and conduction all play a
role in setting the opacity at different depths in the neutron star
atmosphere. In this Appendix, we describe how we calculate each of
these contributions. Prior to H/He burning, the flux in the atmosphere
is carried by radiation and the opacity is set by Thomson scattering,
$\kappa=\kappa_{\rm es}=\sigma_{\rm Th}(1+X)/2m_p$ where $X$ is the
hydrogen mass fraction.  There are corrections to $\kappa_{\rm es}$
due to degeneracy and high temperatures for which we use Paczynski's
(1983) fit to the results of Buchler \& Yueh (1976). This formula is
valid for $\eta\equiv E_F/k_BT\lesssim 4$, where $E_F$ is the electron
Fermi energy excluding the rest mass. We find in our models that when
electron scattering dominates the opacity we are safely in the $ \eta
\lesssim 4 $ regime. 

As the hydrogen and helium are burned, electrons are consumed and the
average nuclear charge of the nuclei increases. This results in
free-free absorption becoming much more important than electron
scattering.  The free-free opacity is given by
\begin{equation}
\kappa_{\rm ff}=0.753 \ {\rm cm^2\over g}\ {\rho_5\over\mu_eT_8^{7/2}}
\ \sum {Z_i^2X_i\over A_i}\ g_{\rm ff}(Z_i,T,n_e),
\end{equation}
(Clayton 1983), where the sum is over all nuclear species and
$\rho_5=\rho/10^5 \ {\rm g \ cm^{-3}}$. The
dimensionless free-free Gaunt
factor $g_{\rm ff}$ takes into account the dependence of the opacity on
Coulomb wavefunction corrections, degeneracy and relativistic effects.
Itoh et al. (1991) have calculated $g_{\rm ff}$ for pure hydrogen, helium,
carbon and oxygen plasmas. So as to implement their work, we 
devised the fitting formula 
\begin{equation}\label{eq:gff}
g_{\rm ff}(Z, T, n_e)=1.16 
\left[{2n_Q\over n_e}\ln(1+e^\eta)\right]
\left[{1-\exp(-2\pi\gamma/\sqrt{\Pi+u})\over 1-\exp(-2\pi\gamma/\sqrt{\Pi})}\right]
\left[1+\left({T_8\over 7.7}\right)^{3/2}\right],
\end{equation}
where $\eta=E_F/k_BT$, $u=10$,
\begin{equation}
{2n_Q\over n_e}\equiv
{2(2\pi m_ek_BT)^{3/2}\over h^3n_e}
=0.08\ T_8^{3/2}\left({\mu_e\over\rho_5}\right),
\end{equation}
\begin{equation}
\gamma^2\equiv{Z^2 (13.6\ {\rm eV})\over k_BT}=1.58\times 10^{-3}\ {Z^2\over T_8},
\end{equation}
and
\begin{equation}
\Pi(\eta)\equiv\left[1+\ln(1+e^\eta)\right]^{2/3}. 
\end{equation}
In the regime $-2.5<\log_{10}\gamma^2<2$, our fitting formula agrees
with their table to better than 10\% for $-6<\eta<5$ and 20\% for
$5<\eta<10$.  In the regime $-4<\log_{10}\gamma^2<-2.5$ and for
$-6<\eta<10$, the agreement is better than 20\% for hydrogen and
within a factor of two for oxygen. This accuracy is adequate for our
application as the $\log_{10}\gamma^2<-2.5$ regime is only relevant
when the composition is mostly hydrogen, in which case electron
scattering dominates. The highest $Z$ element considered by Itoh et
al. (1991) is oxygen ($Z=8$). We assume in our work that our formula,
which includes the important scalings, is applicable to elements with
greater $Z$. To calculate $E_F$, we use the analytic fitting formulae
of Antia (1993) and Chabrier \& Potekhin (1998, eq. [24]).

The three terms in equation (\ref{eq:gff}) are physically
motivated. The first is the degeneracy correction to the electron
velocity.  We follow Cox \& Giuli (1968, \S 16.2), who corrected for
the electron degeneracy by integrating the free-free absorption
cross-section over the Fermi-Dirac distribution.  The second term is
the Elwert factor (Elwert 1939); a correction to the electron
wavefunction normalization due to the Coulomb potential (Pratt \&
Tseng 1975; Lee et al. 1976; Bethe \& Salpeter 1977). The Elwert
factor depends on the ratio of the Coulomb energy to the electron
energy, parameterized by $\gamma^2\equiv Z^2 (13.6\ {\rm eV})/k_BT$,
and the energy of the photon involved in the transition. We write an
average photon energy $u\equiv h\nu/k_BT$, and find that $u=10$ gives
a good fit. The parameter $\Pi(\eta)$ tracks the transition between
$k_BT$ and $E_F$ as the energy scale which sets the electron
velocity. We chose the particular form of the function $\Pi(\eta)$ that
gave the best fit.  The third term includes the effect of relativity
on the Coulomb scattering. Bethe \& Salpeter (1977) show that in the
extreme relativistic limit, the free-free cross section $\sigma_{\rm
ff}$ is proportional to the electron energy $E_e$, whereas in the
non-relativistic case $\sigma_{\rm ff}\propto E_e^{-1/2}$.  The Gaunt
factor rises rapidly at high temperatures to allow for this different
scaling, as found in the numerical calculations of Itoh and coworkers
(Itoh et al. 1985, 1990, 1991; Nakagawa et al. 1987)\footnote{At very
high temperatures and low degeneracy, the opacity increases because of
$e^+$--$e^-$ pair production, increasing the number density of
scatterers. We never encounter this regime in our steady-state models
and so do not include it here in our fitting formula.}. We thus added
a term $\propto T^{3/2}$ at high temperatures, choosing the transition
temperature $T_8=7.7$ which gave the best fit.

The heat transport is dominated by electron conduction once the
material becomes degenerate. Yakovlev \& Urpin (1980, hereafter YU)
wrote the conductivity as
\begin{equation}\label{eq:conduct} 
K={\pi^2 k_B^2 T n_e\over 3 m_* \nu_c}, 
\end{equation}
where $\nu_c=\nu_{ei}+\nu_{ee}$ is the electron collision frequency
and $m_*=m_e+E_F/c^2$ ($E_F$ is the electron Fermi energy not
including the rest mass). We calculate electron-electron collisions
using the fit of Potekhin et al. (1997), but these make a small
contribution to the total collision frequency (typically $\lesssim
5$\%). The conductivity is mainly determined by electron-ion
collisions, for which we have generalized YU's results to an ionic
mixture. The scalings become evident by writing $\nu_{ei}\sim
\sum_i n_i\sigma_i v_F \sim \sum_i n_i v_F Z_i^2e^4\Lambda_{ei}/(p_Fv_F)^2$,
where we sum independently over each ion and $\Lambda_{ei}$ is the
Coulomb logarithm. Since $v_F=p_F/m_*$ and $p_F\propto n_e^{1/3}$,
we find
\begin{equation}\label{eq:cond1}
\nu_{ei}={4e^4m_*\over 3\pi \hbar^3}{\sum_i Y_iZ_i^2\over Y_e}\Lambda_{ei},
\end{equation}
a generalized form of the familiar YU result, where $Y_i\equiv
X_i/A_i$ and we have inserted the correct prefactor as found by
YU. The Coulomb logarithm is
\begin{equation}\label{eq:cond2}
\Lambda_{ei}=\Lambda_{ei}^o-{v_F^2\over 2c^2}; \hskip 20 pt
\Lambda_{ei}^o=\ln(r_{\rm max}/r_{\rm min})
\end{equation}
(YU), where the second term is the relativistic correction to the
cross-section, and $r_{\rm max}$ and $r_{\rm min}$ are the limits of
the integral over impact parameters.  The lower limit is set by the
electron wavelength $r_{\rm min}=\hbar/2p_F$. YU give a good fit for
the upper limit as $r_{\rm max}^2=r_D^2+a^2/6$, where the Debye
screening length $r_D$ for a mixture of ions is $r_D^{-2}=4\pi e^2
(\sum n_i Z_i^2)/k_BT$ (Shu 1991) and $a$ is the average spacing
between ions, defined by $4\pi a^3 \sum n_i/3=1$. If we also define
$\Gamma$ for the mixture as
\begin{equation}\label{eq:cond3}
\Gamma={e^2\over k_BTa}{\sum n_i Z_i^2\over\sum n_i}=
0.49\left({\sum_iY_i Z_i^2\over\sum_iY_i}\right)
{\left(\rho_7\sum_iY_i\right)^{1/3}\over T_8},
\end{equation}
(Hubbard \& Lampe 1969), then we find
\begin{equation}\label{eq:cond4}
\Lambda_{ei}^o=\ln\left[\left({2\pi\over 3}\right)^{1/3}
\left({Y_e\over\sum_iY_i}\right)^{1/3}\left({3\over\Gamma}+
{3\over 2}\right)^{1/2}\right].
\end{equation}
Equation (\ref{eq:cond1}), together with equations (\ref{eq:cond2}),
(\ref{eq:cond3}) and (\ref{eq:cond4}), gives a general expression for
the electron-ion collision frequency for an ion mixture. For the case
of a single species of ion, $\sum_i Y_i Z_i^2/Y_e=Z$ and
$Y_e/\sum_iY_i=Z$ and these equations reduce to the expressions of
YU. Note that the scalings in the mixed case are {\it not} those
obtained by substituting the mean value of Z into the expressions of
YU.

\begin{figure}
\plotone{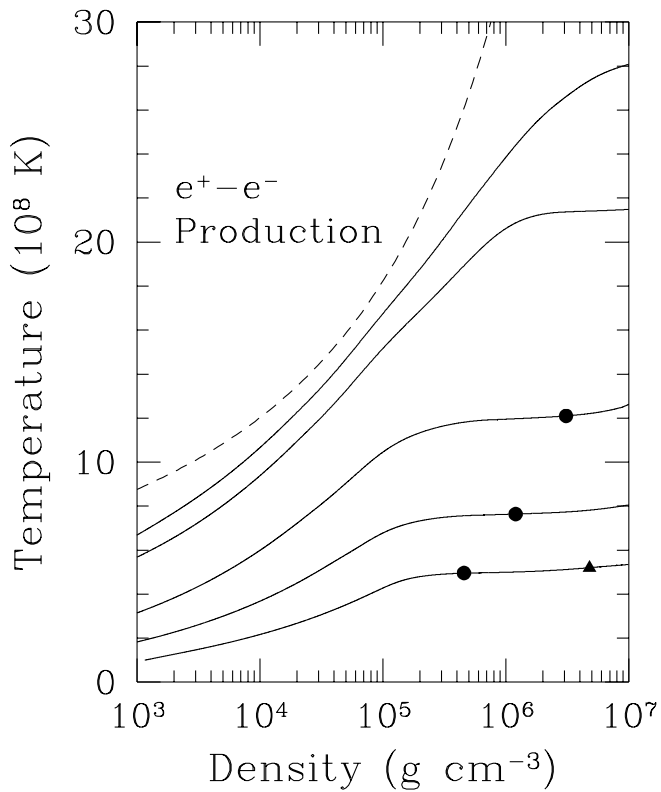}
\caption{
The temperature-density profiles of a rapidly accreting $M=1.4
M_\odot$, $R=10\ {\rm km}$ neutron star atmosphere which is burning in
steady-state. From bottom to top, the curves are for $\dot m/ \dot
m_{\rm Edd}=1, 5, 20, 50$ and $60$, respectively. The number density
of positrons exceeds 10\% of the neutralizing electron density to the
left of the dashed line. For this reason, we do not go above $\dot m/
\dot m_{\rm Edd}=60$. To the left of the filled circle, the opacity is
set by electron scattering ($\kappa_{es}>\kappa_{ff}$); to the right
of this point, free-free opacity is more important
($\kappa_{ff}>\kappa_{es}$). Conduction eventually takes over from
radiation as the heat transport mechanism. For the
$\dot{m}=\dot{m_{Edd}}$ model, we mark with a triangle where
$\kappa_{cond}=\kappa_{ff}$ (for the other models, conduction takes
over at $\rho>10^7\ {\rm g\ cm^{-3}}$).
\label{fig:trho}}
\end{figure}

\begin{figure}
\epsscale{1.0}
\plotone{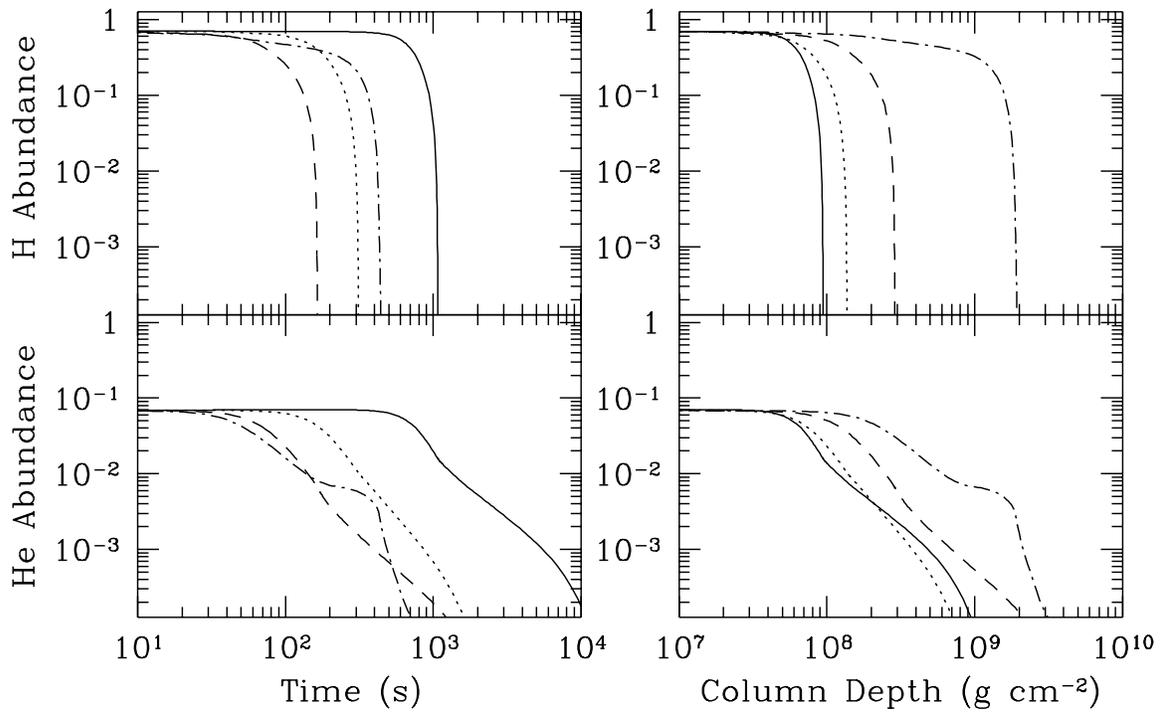}
\caption{Hydrogen and helium abundances as a function of time (left
panel) and column depth (right panel) for accretion rates $\dot m /
\dot m_{\rm Edd}$= 1 (solid line), 5 (dotted line), 20 (dashed line),
and 50 (dot-dashed line).
\label{fig:abut}}
\end{figure}

\begin{figure}
\epsscale{1.0}
\plotone{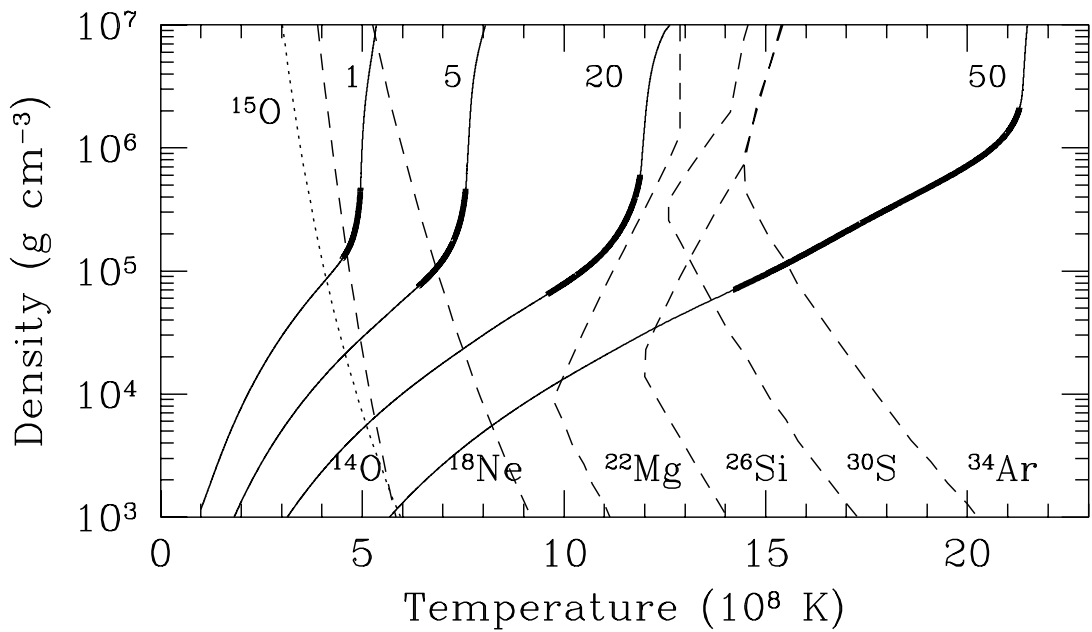}
\caption{
The tracks of a fluid element in the $T$--$\rho$ plane for various
accretion rates. The accretion rate is indicated by the number near
the end of the track in units of $\dot m_{\rm Edd}$. The thick line
segment shows where hydrogen burns from 90\% down to 10\% of its
initial abundance. The dotted line marked ``$^{15}$O'' shows the
conditions where the $^{15}$O($\alpha$,$\gamma$)$^{19}$Ne rate equals
the $^{15}$O $\beta^+$ decay rate. The dashed lines show where the
($\alpha$,p) reaction rates on $^{14}$O, $^{18}$Ne, $^{22}$Mg,
$^{26}$Si, $^{30}$S and $^{34}$Ar equal the other destruction
mechanisms ($\beta^+$ decays and proton captures) on these
isotopes. In the temperature and density region to the right of these
dashed (or dotted) lines, the ($\alpha$,p) (or ($\alpha$,$\gamma$))
reactions dominate the destruction reactions of the respective
isotopes.
\label{tr}}
\end{figure}

In Figure \ref{fig:appkappa}, we show the resulting opacity for pure
$^{56}$Fe in the $\rho$--$T$ plane. The solid lines show where the
dominant opacity source changes from electron scattering to
free-free absorption, and from free-free opacity to conduction,
i.e. where $\kappa_{es}=\kappa_{ff}$ and $\kappa_{ff}=\kappa_{cond}$,
where $\kappa_{cond}=4acT^3/3\rho K$ is the conductive opacity. The
line $\kappa_{ff}=\kappa_{cond}$ agrees well with Figure 2 of
Gudmundsson, Pethick \& Epstein (1983), who used the Los Alamos
opacities for pure $^{56}$Fe. The dotted lines show contours of total
opacity $\kappa$, given by
$\kappa^{-1}=\kappa_{cond}^{-1}+(\kappa_{es}+\kappa_{ff})^{-1}$. To
the left of the long dashed line, significant $e^+$--$e^-$ pair
production occurs. We do not include this in our calculations as we
never encounter this regime in our steady-state models (see Figure
\ref{fig:trho}), thus our opacity calculations are not valid in this
region. We choose iron here as an illustrative example to compare to
previous work. In practice, as we show in this paper, there is a rich
mixture of elements produced by the nuclear burning.

We now compare our opacity calculations to those of previous
workers. The main difference is in the treatment of free-free
opacity. Taam, Woosley \& Lamb (1996) in their time-dependent
simulations used the KEPLER code of Weaver, Zimmerman \& Woosley
(1978) (see also Woosley \& Weaver 1984) which calculates opacities
from the analytic fits of Iben (1975). Iben (1975) provides fitting
formulae to the radiative opacities of Cox \& Stewart (1970a,b) and
the conductive opacities of Hubbard \& Lampe (1969) (for
non-relativistic electrons) and Canuto (1970) (for relativistic
electrons). Fujimoto and coworkers (FHM; Fujimoto et al. 1984; Hanawa
\& Fujimoto 1984, 1986) also used this same fit. Joss (1977), Joss \&
Li (1980) and Ayasli \& Joss (1982) used Iben's fit to the conductive
opacity, but a fit by Stellingwerf (1975) to the radiative opacities
of Cox, King \& Tabor (1973). More recently, Potekhin, Chabrier \&
Yakovlev (1997) used the OPAL opacity library (Rogers, Swenson \&
Iglesias 1996) for their study of cooling neutron stars with accreted
envelopes with pure H, He, C, O and Fe layers. As an example, we have
compared our calculation of the total opacity $\kappa$ with that
obtained using the fit of Iben (1975) for the free-free
contribution. We find that for pure $^{56}$Fe at $\rho>10^{6}\ {\rm g\
cm^{-3}}$ and $T> 5\times 10^8\ {\rm K}$ (the $\rho$ and $T$ regime
for which free-free opacity becomes important in our models), the
total opacity we calculate is 30\% to 50\% larger than that using
Iben's fit.

%----------------------------- REFERENCES -----------------------------

%-------------------------- TABLE -------------------------------
\begin{deluxetable}{lccc}
\tablecaption{
H-burning Duration, Energy Generation, and Amount of Helium Surviving 
H-burning
\label{endpoints}}
\tablewidth{0pt}
\tablehead{
\colhead{$\dot m/\dot m_{\rm Edd}$} &
\colhead{$t_{\rm H}$ (s) \tablenotemark{a}} &
\colhead{$E$ (MeV/nucleon) \tablenotemark{b}} &
\colhead{$Y_r$ (\%) \tablenotemark{c}}
}
\startdata
1& 486 &5.34 &6.7 \nl
5& 185 &5.61 &4.5 \nl
20& 109 &5.65 &2.9 \nl
50& 349 &5.98 &1.4 \nl
\enddata
\tablenotetext{a}{The time for the hydrogen abundance to decrease from 90\% to 10\% of its initial value.}
\tablenotetext{b}{The total energy release from the nuclear burning.}
\tablenotetext{c}{The mass fraction of helium that remains when 99\% of the hydrogen has been burned.}
\end{deluxetable}

%----------------------------- FIGURES

\newpage

\begin{figure}
\epsscale{0.9}
\plotone{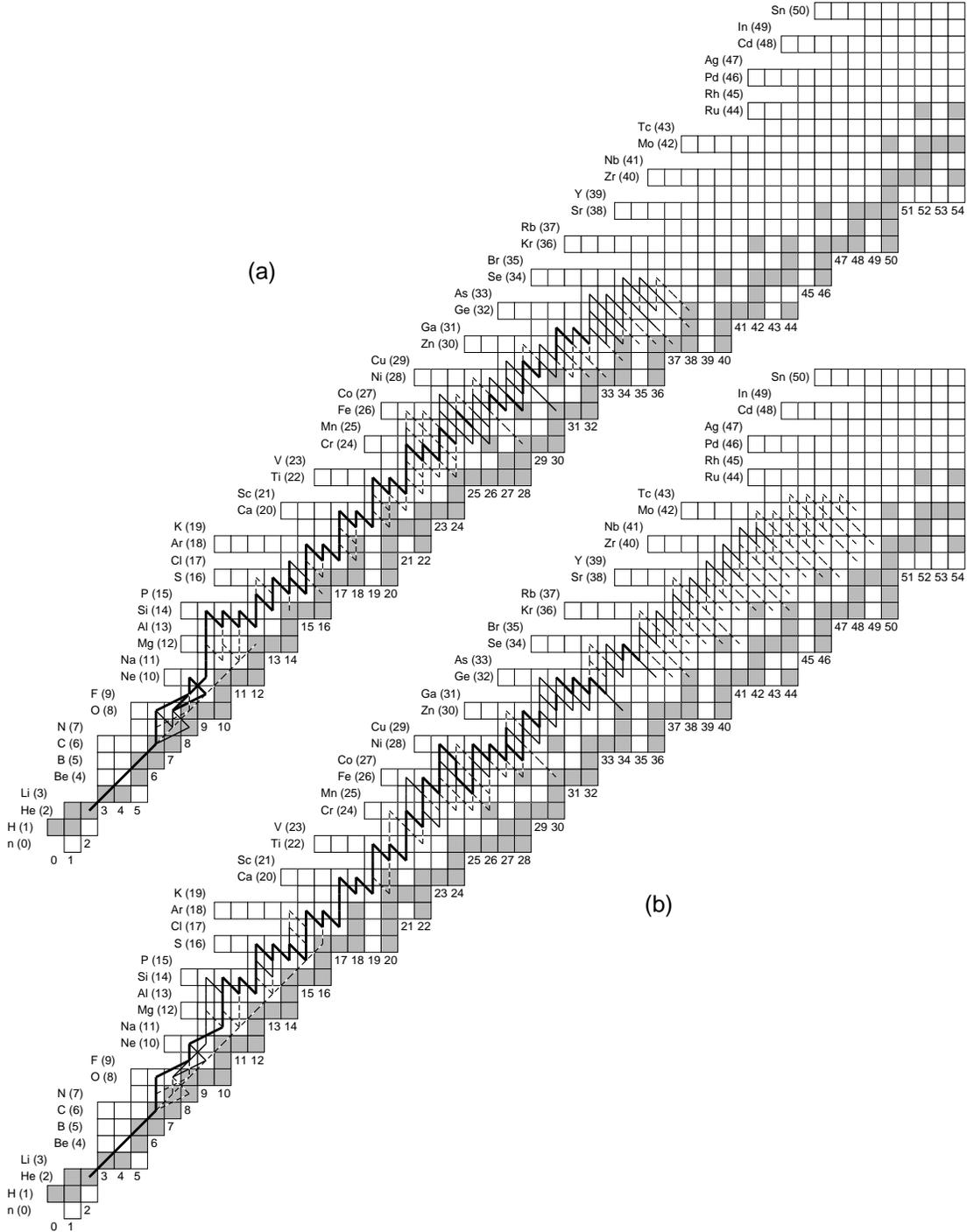}
\caption{
The time-integrated reaction flow for accretion rates of (a) $1\medd$,
(b) $5\medd$, (c) $20\medd$ and (d) $50\medd$. The thickness of each
line indicates the strength of the reaction flow relative to the
3$\alpha$-reaction: more than 50\% flow (thick solid line), 10\%--50\%
flow (thin solid line), and 1\%--10\% flow (dashed line). The
3$\alpha$ reaction is a useful normalization reaction since
essentially all the reaction flow passes through that reaction for all
accretion rates. Each square stands for a proton stable nucleus,
filled squares are stable nuclei and ``P'' indicates a
p-nucleus. Nuclei on the neutron rich side of stability have been
omitted.
\label{flow}}
\end{figure}

\begin{figure}
\epsscale{0.9}
\figurenum{\ref{flow}}
\plotone{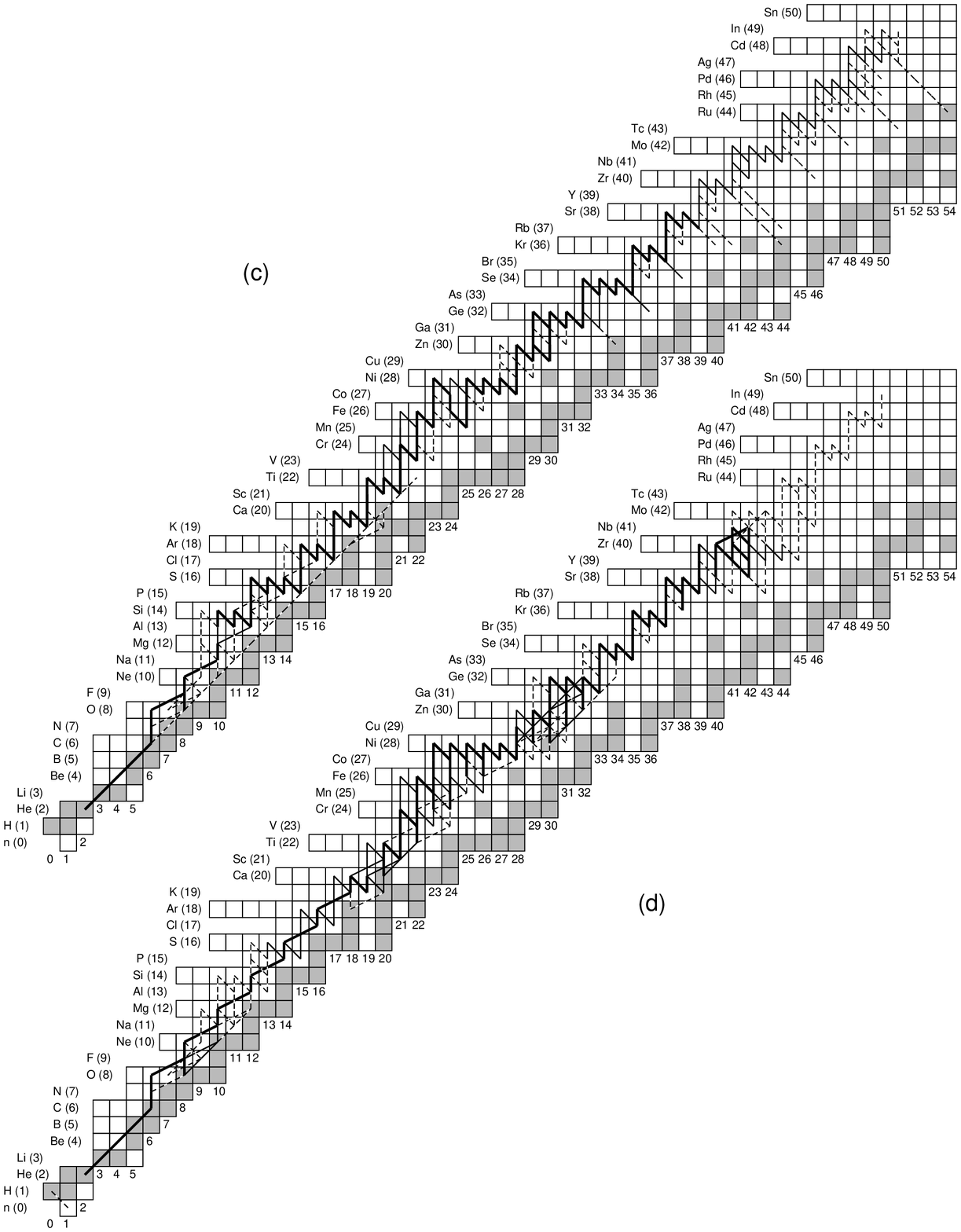}
\caption{Continued.}
\end{figure}

\begin{figure}
\epsscale{0.6}
\plotone{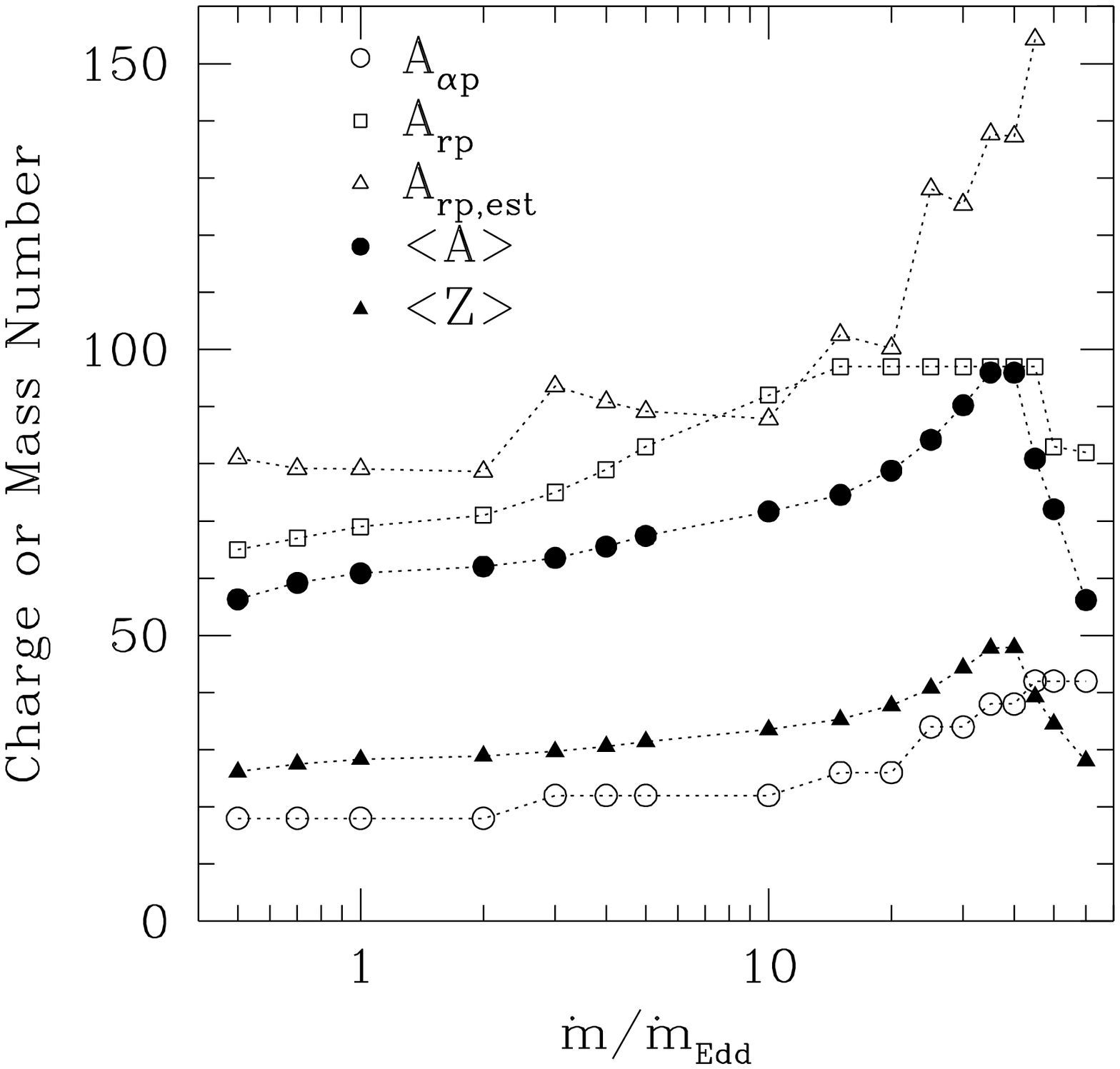}
\caption{
The mass number of the last isotope produced in the $\alpha$p process
($A_{\alpha p}$, open circles) and in the rp process ($A_{\rm rp}$,
open squares) as a function of the local accretion rate
$\dot{m}/\dot{m}_{\rm Edd}$.  The last isotope produced is defined as
the isotope where the time integrated reaction flow of the respective
process drops below 10\% of its maximum. The open triangles show
$A_{\rm rp, est}$, the estimate of $A_{\rm rp}$ from $A_{\alpha p}$
using equation (\ref{eq:arp}). We also show the average mass
number ($<A>$, filled circles) and the average charge number ($<Z>$,
filled triangles) of the final composition (excluding nuclei with
$A\le12$). We connect the data points with straight lines to guide the
eye.
\label{mrange}}
\end{figure}

\begin{figure}
\epsscale{1.0}
\plotone{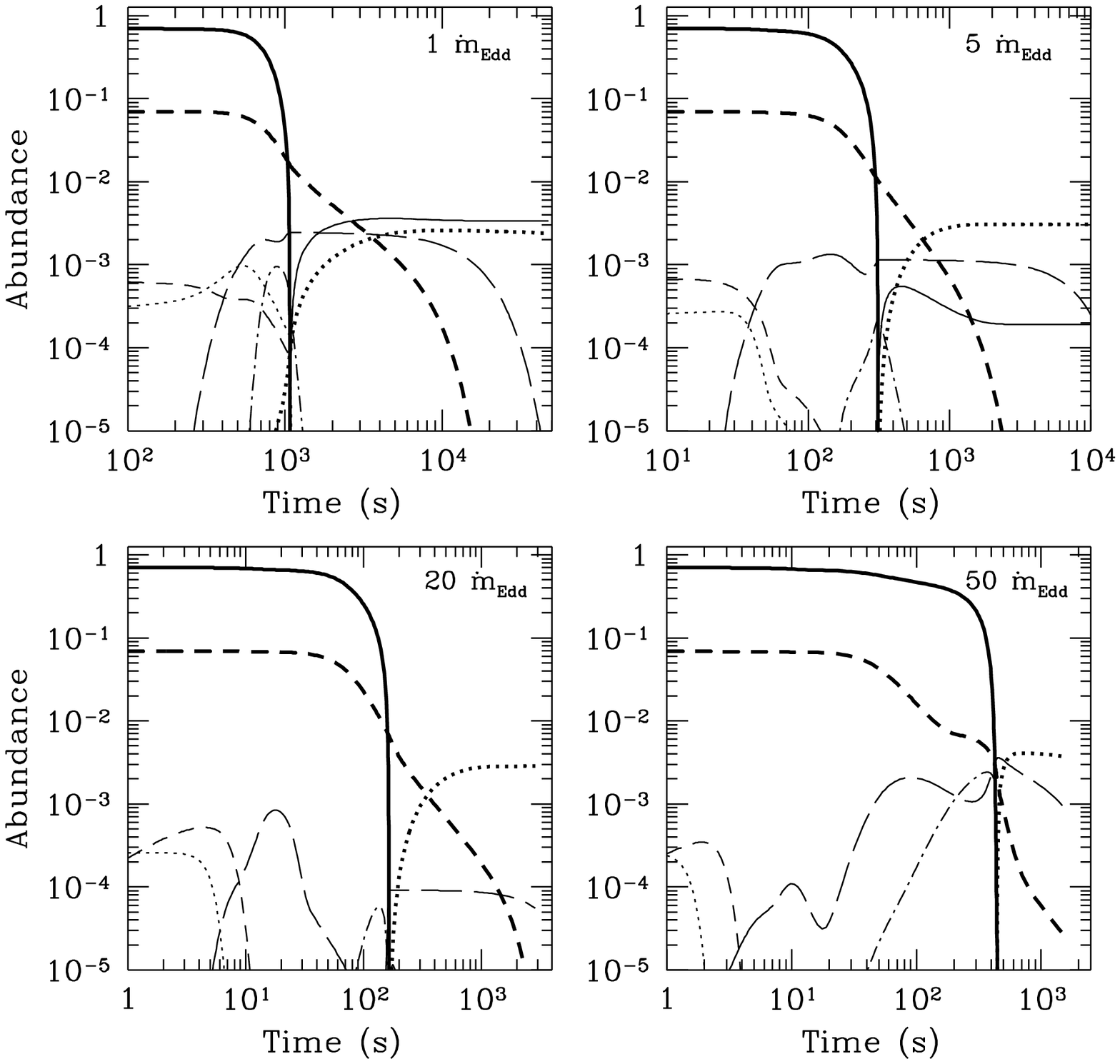}
\caption{The abundances of various isotopes as a function of time 
for accretion rates $\dot m / \dot m_{\rm Edd}$=1, 5, 20, and 50. In
each figure, we show abundances of $^{1}$H (bold solid line),
$^{4}$He (bold dashed line), $^{12}$C (solid line), $^{14}$O (dotted
line), $^{15}$O (short dashed line) and $^{56}$Ni (long dashed
line). The dot-dashed line shows the last isotope produced by the rp
process in each case. This is $^{67}$As, $^{80}$Y, $^{96}$Cd and
$^{80}$Zr for $\dot m / \dot m_{\rm Edd}$=1, 5, 20, and 50
respectively. The thick dotted line shows an abundant element in each
case at the end of our calculations. This is $^{67}$Ga, $^{64}$Zn,
$^{68}$Ge and $^{80}$Sr for $\dot m / \dot m_{\rm Edd}$=1, 5, 20, and
50 respectively.
\label{abut2}}
\end{figure}

\begin{figure}
\epsscale{0.6}
\plotone{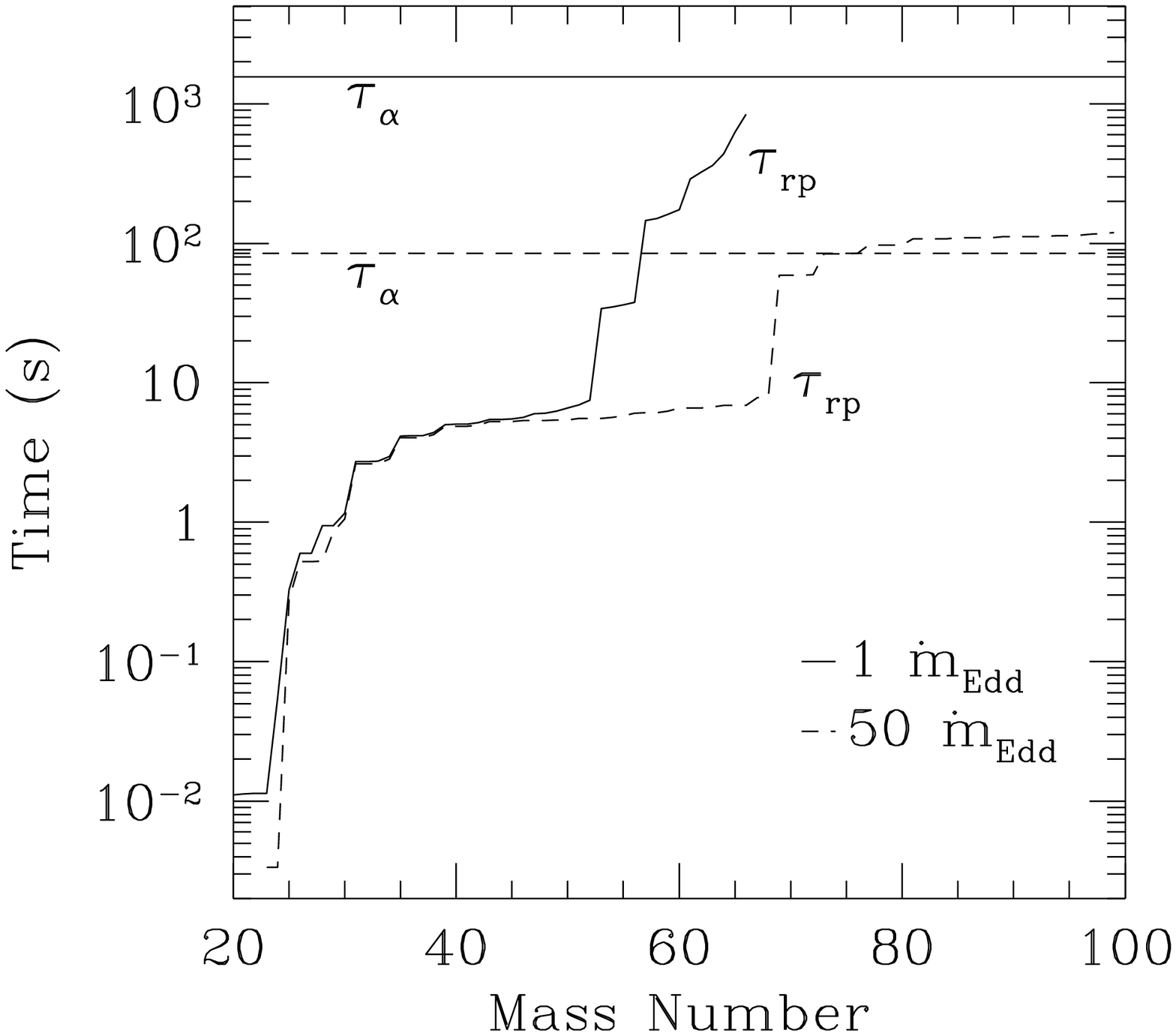}
\caption{
The time it takes the rp process to reach a given mass number
$\tau_{\rm rp}$ compared to the timescale for helium burning via the
3$\alpha$-reaction $\tau_\alpha$. We show results for two different
accretion rates: $\dot m=\dot m_{\rm Edd}$ (solid lines) and
$\dot{m}=50\dot{m}_{\rm Edd}$ (dashed lines). For $\dot m=\medd$,
$\tau_{\rm rp}\ll\tau_\alpha$, whereas for $\dot m=50\medd$ the
timescales become comparable for $A\gtrsim 70$. This explains why the
amount of helium remaining when hydrogen is exhausted ($Y_r$) is less
for higher accretion rates (Table \ref{endpoints}).
\label{tau}}
\end{figure}

\begin{figure}
\epsscale{1.0}
\plotone{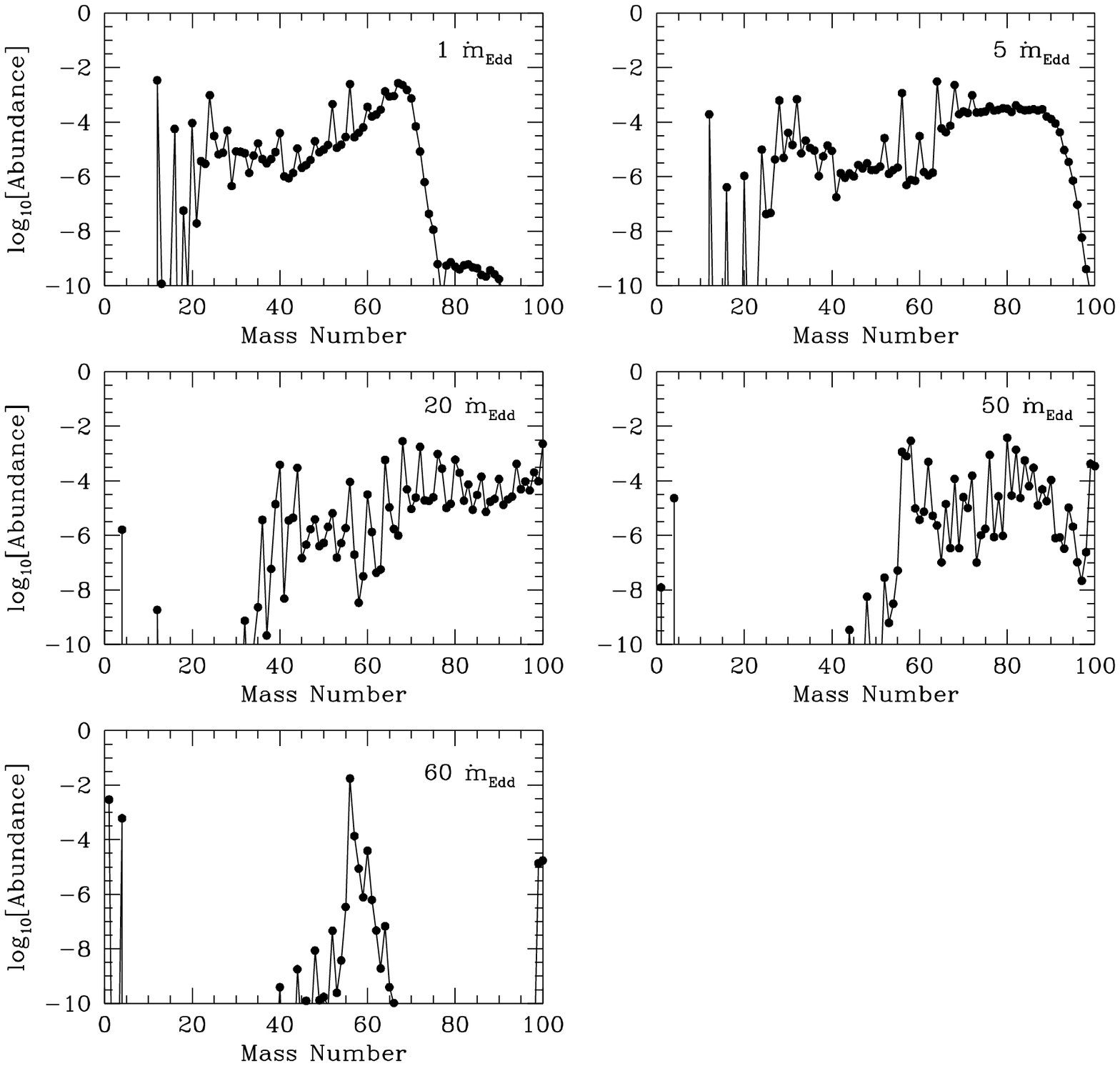}
\caption{The distribution of the final abundances for
accretion rates of $\dot m / \dot m_{\rm Edd}$=1, 5, 20, 50 and 60.
The final isotopic abundances have been summed for each mass number
$A$ and the resulting abundances are shown as functions of
$A$. \label{abu}}
\end{figure}

\begin{figure}
\epsscale{0.7}
\plotone{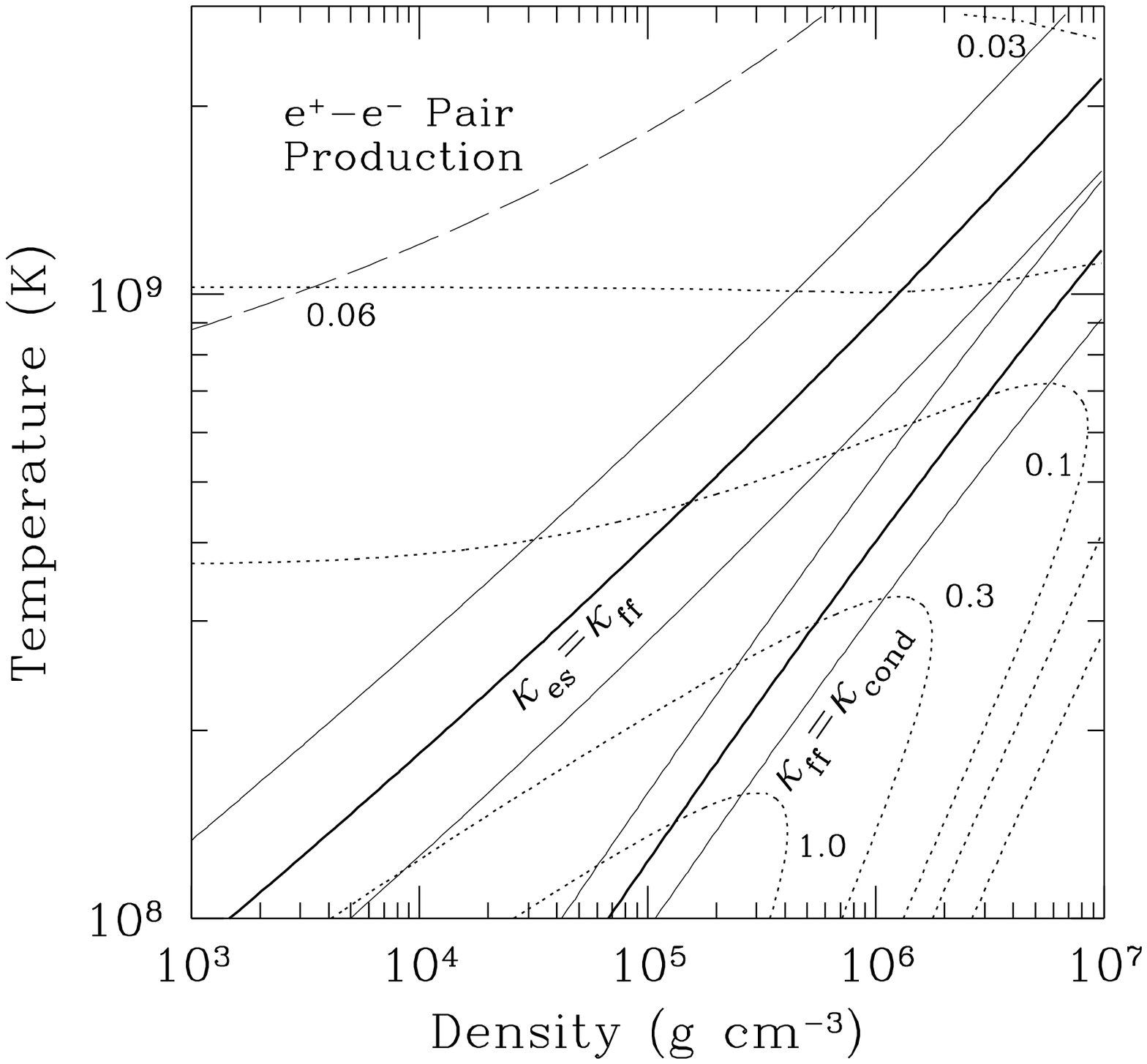}
\caption{
Contours of opacity for pure $^{56}$Fe in the $\rho$--$T$ plane
(dotted lines). Each contour is labelled with the value of opacity
$\kappa$. The heavy solid lines show where the contributions due to
conduction ($\kappa_{cond}$) or electron scattering ($\kappa_{es}$)
are equal to the free-free opacity $\kappa_{ff}$. The light solid
lines show where the ratios $\kappa_{ff}/\kappa_{es}$ or
$\kappa_{ff}/\kappa_{cond}$ are equal to 0.25 or 4. The number density
of positrons exceeds 10\% of the neutralizing electron density to the
left of the dashed line: our opacity calculations are not valid in
this region (we do not encounter this regime in our models).
\label{fig:appkappa}}
\end{figure}

%  ------------------------------------------------------------------

\end{document}